\documentclass[lettersize,journal]{IEEEtran}
\usepackage{amsmath,amsfonts}
\usepackage{array}
\usepackage{textcomp}
\usepackage{stfloats}
\usepackage{url}
\usepackage{verbatim}
\usepackage{graphicx}
\usepackage{cite}
\usepackage{booktabs}
\usepackage{xcolor}

\usepackage{multirow}
\usepackage{diagbox}
\usepackage{makecell}
\usepackage{array}

\usepackage{algpseudocode}
\usepackage[ruled,vlined]{algorithm2e}

\usepackage{tablefootnote}
\usepackage{threeparttable}

\usepackage{subcaption}

\begin{document}

\title{IDEA: An Inverse Domain Expert Adaptation Based Active DNN IP Protection Method}

\author{
        {
                Chaohui~Xu,
                Qi~Cui,~\IEEEmembership{Member,~IEEE,}
                Jinxin~Dong,
                Weiyang~He,
                and~Chip-Hong~Chang,~\IEEEmembership{Fellow,~IEEE}
        }
        \thanks{This research is supported by the National Research Foundation, Singapore, and Cyber Security Agency of Singapore under its National Cybersecurity Research \& Development Programme (Development of Secured Components \& Systems in Emerging Technologies through Hardware \& Software Evaluation \texttt{<} NRF-NCR25-DeSNTU-0001 \texttt{>}). Any opinions, findings and conclusions or recommendations expressed in this material are those of the author(s) and do not reflect the view of National Research Foundation, Singapore and Cyber Security Agency of Singapore.}
        \thanks{C. Xu, J. Dong, and W. He are with the School of Electrical and Electronic Engineering, Nanyang Technological University, Singapore 639798. Q. Cui is with the Engineering Research Center of Digital Forensics, Ministry of Education, School of Computer Science, Nanjing University of Information Science and Technology, Nanjing, China 210044. C. H. Chang is with the School of Electrical and Electronic Engineering and National Integrated Centre for Evaluation (NiCE), Nanyang Technological University, Singapore 639798. (Email: \{chaohui001, echchang\}@e.ntu.edu.sg). Corresponding author: C. H. Chang.}
}

\markboth{Journal of \LaTeX\ Class Files,~Vol.~14, No.~8, August~2021}%
{Shell \MakeLowercase{\textit{et al.}}: A Sample Article Using IEEEtran.cls for IEEE Journals}


\maketitle

\begin{abstract}

Illegitimate reproduction, distribution and derivation of Deep Neural Network (DNN) models can inflict economic loss, reputation damage and even privacy infringement. Passive DNN intellectual property (IP) protection methods such as watermarking and fingerprinting attempt to prove the ownership upon IP violation, but they are often too late to stop catastrophic damage of IP abuse and too feeble against strong adversaries. In this paper, we propose IDEA, an Inverse Domain Expert Adaptation based proactive DNN IP protection method featuring active authorization and source traceability. IDEA generalizes active authorization as an inverse problem of domain adaptation. The multi-adaptive optimization is solved by a mixture-of-experts model with one real and two fake experts. The real expert is re-optimized from the source model to correctly classify authorized images embedded with the correct user key. The two fake experts are separately trained to output random prediction on test images without or with incorrect user key embedded, respectively. The constructed MoE model is knowledge distilled into a unified protected model to avoid functionality leakage. IDEA not only prevents unauthorized users without the valid key to access the functional model, but also enable secure licensee verification and reliable ownership verification. We extensively evaluate IDEA on four datasets and four DNN models to demonstrate its stealth, effectiveness, fidelity, uniqueness, and robustness against various attacks.

\end{abstract}


\section{Introduction}

The past decade has witnessed proliferate applications of DNNs in almost every sector of business. Building a high-performance DNN requires significant data collection and labeling effort, computing resources, and expert knowledge. These painstakingly trained models are alluring targets of piracy and misappropriation. When a trained DNN is directly distributed to the end users, the internal structure and parameters of the deployed model can be easily duplicated by rival adversaries or dishonest consumers with full access to the model. Even if the trained model is deployed on the cloud to remote endpoint users for online inference through application program interface (API), recent studies~\cite{tramer2016stealing,orekondy2019knockoff,kariyappa2021maze,rakin2022deepsteal,sanyal2022towards} demonstrated that a surrogate model can be trained to achieve comparable performance as the source model at a substantially reduced cost than designing the model from scratch.

In view of the rampant DNN intellectual property (IP) theft, various protection methods have been proposed in recent years. The approaches can be categorized into passive and active protection methods. Passive protection methods, such as watermarking~\cite{uchida2017embedding,le2020adversarial,adi2018turning,szyller2021dawn,chen2019deepmarks,wang2021riga,namba2019robust} and fingerprinting~\cite{jia2021proof,cao2021ipguard,li2021modeldiff,zheng2022dnn}, aim at detecting the ownership of the deployed DNN upon IP infringement, whereas active protection methods~\cite{chakraborty2020hardware,pyone2020training,fan2021deepipr,lin2020chaotic,xue2023ssat,zhou2023nnsplitter,li2024securenet} aim at preventing IP violation proactively. This work mainly focuses on active protection which safeguards the target model by locking its functionality to unauthorized usage. That is, the protected model performs normally without substantially degraded inference performance only when the correct key is present.

Research on active protection methods are in the nascent state of development. Existing active protection solutions face notable challenges, including dependence on hardware~\cite{chakraborty2020hardware,zhou2023nnsplitter}, low fidelity~\cite{pyone2020training}, and limited robustness~\cite{xue2023ssat,pyone2020training,fan2021deepipr,zhang2020passport}. Notably, some approaches~\cite{xue2023ssat,li2024securenet,ren2022protecting,tang2020deep} achieve active protection through multi-task optimization, aiming for the model to perform well only on authorized samples. However, these techniques mainly alter the decision boundaries of the last and/or penultimate layers. Consequently, the feature distribution in the shallow layers remains almost identical for both authorized and unauthorized samples, making the protected model susceptible to fine-tuning attacks.

In this paper, we propose IDEA -- an \textbf{I}nverse \textbf{D}omain \textbf{E}xpert \textbf{A}daptation based multi-user active DNN IP protection framework. Unlike existing methods, IDEA treats the active authorization IP protection as an inverse problem of domain adaptation, and solves it by training a mixture of experts (MoE) to minimize the mutual information (MI) between authorized source and unauthorized target domains. Encoded images are produced by embedding a binary string (referred to as key) using a pretrained SteganoGAN~\cite{zhang2019steganogan}. The encoder of the SteganoGAN is distributed to all legitimate users, while the decoder is kept private by the model owner. To construct the MoE model, a real expert is fine-tuned from the unprotected model to produce accurate prediction on authorized images (encoded with the correct key), and two fake experts are trained to intentionally degrade the prediction accuracy on clean images (unencoded) and noise images (encoded with an invalid key), respectively. The three independently trained experts are inseparably fused into a coherent protected model by knowledge distillation across several selected hidden layers from shallow to deep. Consequently, only the authorized user who holds the correct key can generate authorized images that the protected model can process normally.

Our contributions are as follows.

\begin{itemize}
  \item We systematically define the requirements for multi-user active DNN IP protection, and propose IDEA which constructs a MoE model with one real expert and two fake experts, and further distill it into a coherent protected model.
  \item We also offers two verification mechanisms: 1) Licensee verification. Authorized users can securely prove their legitimacy by simply providing encoded test images without having to divulge their secret keys; 2) Ownership verification. Even if the model has been misappropriated, the owner can still conduct black-box queries on the suspected model to claim ownership and trace the culprit responsible for the model leakage.
  \item We conduct extensive evaluations across four datasets and four model architectures to demonstrates the superior performance of IDEA in terms of stealth, effectiveness, fidelity, and uniqueness. Its superiority over existing active protection methods and robustness against a wide range of attacks have also been validated.
\end{itemize}

The rest of the paper is organized as follows. Related works are reviewed in Section~\uppercase\expandafter{\romannumeral2}. Our threat model and problem formulation, background knowledge on SteganoGAN and domain expected risk bound are presented in Section~\uppercase\expandafter{\romannumeral3}. The proposed IDEA is elaborated in Section~\uppercase\expandafter{\romannumeral4}. Section~\uppercase\expandafter{\romannumeral5} presents the experimental results and comparison, followed by the latent representations visualization and security analysis in Sections~\uppercase\expandafter{\romannumeral6} and~\uppercase\expandafter{\romannumeral7}, respectively. Limitations of IDEA are discussed in Section~\uppercase\expandafter{\romannumeral8}. The paper is concluded in Section~\uppercase\expandafter{\romannumeral9}.

\section{Related Works}
\label{sec:related_works}

\subsection{Passive Protection}

DNN watermarking is a passive means of IP rights protection by concealing the copyright information into the target model. In the event of IP piracy, the embedded information can be extracted to prove the ownership. The first DNN watermarking method~\cite{uchida2017embedding} embeds secret information into the model's weights by including an additional regularization loss during training and constraining the biases of the embedded hidden layers to follow a particular distribution. To enhance the watermark robustness against colluded modifications by malicious users, DeepMark~\cite{chen2019deepmarks} assigns each user a unique fingerprint using anti-collusion codebook and retrains the model with a fingerprint-specific regularization loss. To avoid accuracy degradation in black-box watermarking, RIGA ~\cite{wang2021riga} uses adversarial training to increase the covertness and robustness against watermark removal but white-box access is needed for watermark verification. To facilitate ownership verification by querying the model remotely via the API, adversarial examples~\cite{le2020adversarial} are used to tweak the decision boundaries for watermark embedding. For better transferability across various model architectures, the model is watermarked by backdooring~\cite{adi2018turning}. The watermarking framework is generalized in~\cite{zhang2018protecting} to support both black-box and white-box accesses by leveraging meaningful content, unrelated content, and meaningless noise for backdoor trigger generation.

In contrast, DNN fingerprinting methods extract unique intrinsic characteristics of the model for ownership proof without modifying the pretrained model. PoL~\cite{jia2021proof} creates a unique fingerprint from specific details of the training process. IPGuard~\cite{cao2021ipguard} utilizes adversarial examples while ModelDiff~\cite{li2021modeldiff} exploits the model's decision distance vectors to characterize and identify the decision boundary for unique fingerprint construction. Both methods allow remote black-box ownership verification through the APIs. More recently, a transfer learnt resilient fingerprint~\cite{zheng2022dnn} is proposed by projecting selected weights from the front layers of the victim model to the random space generated by the owner's notary identity.

Apart from image classification tasks, passive protection methods for generative models such as generative adversarial networks (GANs)~\cite{ong2021protecting,fei2023robust,qiao2023novel} and large language models (LLMs)~\cite{zhang2024reef,pasquini2024llmmap}, have also been explored in recent years.

\subsection{Active Protection}

Active protection methods control the model usage by allowing the full functionality of the model to be accessible by only authorized users. Our method falls into this category.

HPNN~\cite{chakraborty2020hardware} embeds the encoded model into a trustworthy hardware device. NNSplitter~\cite{zhou2023nnsplitter} utilizes a reinforcement learning algorithm to identify a few important filters and obfuscates them to lock the target model's normal functionality. This method relies on trusted execution environment to secure the model inference, which is not flexible and scalable without inference time overhead. In~\cite{pyone2020training}, the model is trained on pixel-shuffled samples to attain optimal inference performance exclusively on test samples that have been shuffled with the same secret key. This method exhibits low fidelity as its pixel shuffling operation distorts the semantic integrity of test images.

M-LOCK~\cite{ren2022protecting}, SSAT~\cite{xue2023ssat}, and SecureNet~\cite{li2024securenet} train the model with correctly labeled trigger-embedded training samples and wrongly labeled clean training samples, such that the protected model can only accurately classify images with the pre-defined backdoor trigger. Similarly, DSN~\cite{tang2020deep} pastes a binary pattern trigger on training images, and utilizes a gradient reversal layer~\cite{ganin2015unsupervised} to train the protected model to behave differently on images with and without the trigger. NTL~\cite{wang2021non} restricts the model's generalization ability to a certain domain to realize active authorization. However, these methods focus on only modifying the decision boundaries of the last and/or penultimate layers, while the feature representations in the shallow layers still leaks the functionality of the protected model.

DeepIPR~\cite{fan2021deepipr}, initially designed for DNN watermarking, can also be employed for active protection. By replacing selected normalization layers of the DNN with passport layers, the protected model performs normally only when the valid passport is presented together with the input. Following DeepIPR, several advanced passport-based protection methods~\cite{zhang2020passport,liu2023trapdoor,cui2024steganographic} are further proposed with improved fidelity, robustness, and flexibility. Unfortunately, the passport layers can be easily located and attacked. Chen et al.~\cite{chen2023effective} proposed an effective ambiguity attack against passport-based protection methods by substituting the located passport layers.

\subsection{Domain Adaption}

Domain adaptation aims to improve the generalizability of a DNN model from the source domain to other related target domains with access to the target training data. Many existing domain adaptation approaches align domain distributions by minimizing the measured distance between domains, such as maximum mean discrepancy (MMD), correlation alignment (CORAL)~\cite{sun2016deep}, and contrastive domain discrepancy (CDD)~\cite{kang2019contrastive}. Some approaches utilize neural networks, like autoencoders or adversarial-based networks~\cite{ajakan2014domain}, to diminish the domain gap by decreasing the discrepancy between feature representations in the hidden layers.

\section{Preliminaries}

\subsection{Threat Model}

The model owner who invests significant effort and resources into training a high-performance DNN model can merchandize its model either as an online service or a distributed IP. The online revenue model allows multiple users to register and pay for inference services either on subscription or per-use basis through API calls to the remotely deployed model. It protects the model owner from the risk of IP theft and model misappropriation, but users may have concern about exposing sensitive or private information when directly submitting test samples to the model owner for predictions. Another revenue model is to sell distributed instances of a DNN model to users at a loyalty fee or blanket IP release price for conditional (e.g., non-exclusive license, non-redistribution, lawful use, etc.) unlimited use on users' local devices for offline predictions. This IP distribution revenue model poses a severe threat to the model IP due to difficulty in enforcement of certain sale agreements. The model's architecture and weights become fully accessible to authorized users. A malicious attacker could steal the deployed model from an authorized user and use it illegally without paying or abuse it for AI scams and frauds. Even more concerning is the possibility of a malicious authorized user who, after purchasing the model once, illegally deploys it on the cloud to provide API-based predictions for profit. Such misuse results in substantial financial losses for the owner.

\textbf{Owner's knowledge and capabilities:} The owner is assumed to have full knowledge and control over all components and training pipelines of the model prior to its distribution and deployment. This enables the owner to modify the training dataset and optimization process as needed to generate multiple protected versions of the model.

\textbf{Owner's goals:} Each protected model should function normally only when the correct key is provided. This ensures that authorized users who have purchased the model can use it without performance degradation, while attackers without a key or other authorized users with mismatched keys cannot use it to obtain accurate inference. Authorized users should be able to prove their right of use by presenting a few, or even a single, encoded image to the owner without revealing the secret key. In the event of illegal redeployment, the owner must be able to assert ownership of the suspected model and identify the authorized user responsible for the model's leakage or misuse. To counteract potential modifications by attackers attempting to evade tracing, the protection method must be robust against various attacks, such as fine-tuning, model pruning, transfer learning, and reverse engineering.

\subsection{Problem Formulation} \label{sec:problem_formulation}

We formulate the active authorization as an inverse problem of domain adaptation. Our goal is to maintain a high inference accuracy of the protected model on the authorized (source) domain, where the input data are embedded with the correct key, while weakening its generalizability on unauthorized (target) domains with missing or incorrect keys. Given a pretrained encoder $\mathcal{E}(\cdot, \cdot)$ and a binary key $k$ which encodes the identity information of both the owner and a specific user (e.g., ``Alice-Bob''), the following three domains of input samples can be defined.
\begingroup
\begin{align}
        \mathcal{B}   &=\left\{(x, y) \mid x \sim \mathcal{P}_X^\mathcal{B}, y \sim \mathcal{P}_Y^\mathcal{B}\right\}, \\
        \mathcal{A}   &=\left\{(x^+, y) \mid x^+:=\mathcal{E}(x, k), x^+ \sim \mathcal{P}_X^{\mathcal{A}}, y \sim \mathcal{P}_Y^{\mathcal{A}}\right\}, \\
        \mathcal{N}   &=\left\{(x^-, y) \mid x^-:=\mathcal{E}(x, k^{\ast}), x^- \sim \mathcal{P}_X^{\mathcal{N}}, y \sim \mathcal{P}_Y^{\mathcal{N}}\right\}.
\end{align}
\endgroup

The benign domain $\mathcal{B}$ consists of the original samples, where $\mathcal{P}_X^\mathcal{B}$ and $\mathcal{P}_Y^\mathcal{B}$ denote the distributions of benign images and labels, respectively. The authorized domain $\mathcal{A}$ contains authorized samples that are all embedded with the correct key $k$. The noise domain\footnote{It should be emphasized that a noise image in IDEA context refers to an unauthorized test image with a wrong embedded key. It should not be misinterpreted as Gaussian noise or visible perturbations as the noise image may not necessarily look distorted or noisy since the wrong key is stegnographically embedded.} $\mathcal{N}$ contains samples that are each embedded with a different random wrong key $k^{\ast} \neq k$.

We consider an image classifier $f$ trained on the benign training dataset $D_\text{tr}=\{(x_i, y_i)\}_{i=1}^{N_\text{tr}}$. Instead of directly releasing the unprotected model $f$ to $M$ paying customers, the model owner further generates $M$ different protected models $\{f^\prime_1,f^\prime_2,\cdots,f^\prime_M\}$ from $f$ by embedding into each instance one of the $M$ unique user-specific keys $\mathcal{K} = \{k_1, k_2,\cdots,k_M\}$. Each protected model has special prediction behaviors on input samples from different domains. For brevity, the user index is omitted, and only a single protected model and corresponding user key are presented below.

\textbf{Definition 1} (Key-based Active Authorization). \textit{Given the benign domain $\mathcal{B}$, a specific binary key $k$, and the corresponding authorized and noise domains $\mathcal{A}$ and $\mathcal{N}$, respectively. A protected model $f^\prime$ is an encoded instance which performs well only on $\mathcal{A}$ but predicts randomized outputs on $\mathcal{B}$ and $\mathcal{N}$ like an untrained model. The protected model fulfills the following three properties:
\begingroup
\begin{align}
        & \left| \operatorname{Acc}\{f, \mathcal{B}\} - \operatorname{Acc}\{f^\prime, \mathcal{A}\} \right| < \epsilon_{1}, \label{eq:fidelity}\\
        & \operatorname{Acc}\{f, \mathcal{B}\} - \operatorname{Acc}\{f^\prime, \mathcal{B}\} > \epsilon_{2},\label{eq:effectiveness}\\
        & \operatorname{Acc}\{f, \mathcal{B}\} - \operatorname{Acc}\{f^\prime, \mathcal{N}\} > \epsilon_{2},\label{eq:uniqueness}
\end{align}
\endgroup
where $\operatorname{Acc}\{\cdot, \cdot\}$ evaluates the provided model's performance on a specific domain or test dataset. $\epsilon_{1}$ and $\epsilon_{2}$ denote two positive fractions that set the accuracy preservation and degradation thresholds, respectively.}

Eq.~\eqref{eq:fidelity} specifies the \textbf{fidelity} requirement on authorized access, which requires the protected model $f^\prime$ to guarantee performance on images encoded with the correct key $k$. Eq.~\eqref{eq:effectiveness} stipulates the \textbf{effectiveness} against unauthorized access, which requires the accuracy of $f^\prime$ to fall below that of the original model by a substantial margin in the absence of $k$. Eq.~\eqref{eq:uniqueness} specifies the \textbf{uniqueness} criterion, which calls for a unique key to affiliate with each encoded model instance. Mismatched model-key pairs will result in a substantial accuracy degradation from its source model. Each model-key pair (i.e., $\{f^\prime, k\}$) along with the unified encoder $\mathcal{E}$ are distributed to the corresponding legal user.

\subsection{SteganoGAN} \label{sec:SteganoGAN}

SteganoGAN~\cite{zhang2019steganogan} is a lightweight steganographic network for key encoding and decoding. It consists of three sub-networks: an encoder $\mathcal{E}$ which imperceptibly embeds binary messages into input images; a decoder $\mathcal{D}$ which extracts the embedded messages from encoded images; and a critic $\mathcal{C}$ reduces distortion in the encoded images. SteganoGAN is employed in IDEA to hide and retrieve secret message imperceptibly into the input image due to its simple architecture, which reduces the computational overhead for key embedding and extraction.

\subsection{Domain Expected Risk Bound} \label{sec:domain_expected_risk_bound}

Given the source domain $\mathcal{S}$ and a target domain $\mathcal{T}$ of a DNN classifier $f$, the expected risk on the source and target domains can be represented as $\mathcal{R}_{\mathcal{S}}(f)=\mathbb{E}_{(x, y) \sim \mathcal{S}}\left[\mathbb{I}\{f(x) \neq y\}\right]$ and $\mathcal{R}_{\mathcal{T}}(f)=\mathbb{E}_{(\hat{x}, y) \sim \mathcal{T}}\left[\mathbb{I}\{f(\hat{x}) \neq y\}\right]$, respectively. The expected risk on the target domain is upper bounded by the latent features as follows~\cite{zhao2021domain}:
\begingroup
\begin{equation}
        \mathcal{R}_{\mathcal{T}}(f) \leq \mathcal{R}_{\mathcal{S}}(f) - 4 I(z;\hat{z}) + 4 H(Y)+\frac{1}{2} d_{\mathcal{H} \Delta \mathcal{H}}\left(p(z), p(\hat{z})\right),
\end{equation}
\endgroup 
where $z$ and $\hat{z}$ represent the latent features extracted independently from the same hidden layer on the source and target domains, respectively. $I(z;\hat{z})$ represents the MI between the latent features. $H(Y)$ is the entropy of label distribution. $d_{\mathcal{H} \Delta \mathcal{H}}(p(z), p(\hat{z}))$ is the $\mathcal{H} \Delta \mathcal{H}$-divergence~\cite{zhao2021domain} between the source and target domain feature marginal distributions. 

Depredating the generalization performance on $\mathcal{T}$ requires the expected risk $\mathcal{R}_{\mathcal{T}}(f)$ to increase. For domain adaptation~\cite{zhao2021domain,guo2024domain}, it is sufficient to increase the expected risk on $\mathcal{T}$ by only minimizing the upper bound of MI between $z$ and $\hat{z}$, $I(z;\hat{z})$. This upper bound can be determined by the Contrastive Log-ratio Upper Bound (CLUB)~\cite{cheng2020club} as follows:
\begingroup
\begin{equation}
        \begin{aligned}
                I(z;\hat{z}) & = \mathbb{E}_{p(z,\hat{z})}\left[\log \frac{p(z,\hat{z})}{p(z)p(\hat{z})}\right] \leq I_{\text{CLUB}}(z;\hat{z}) \\
                & = \mathbb{E}_{p(z,\hat{z})}\left[\log p(\hat{z}|z)\right]-\mathbb{E}_{p(z) p(\hat{z})}\left[\log p(\hat{z}|z)\right],
        \end{aligned}
\end{equation}
\endgroup
where $p(z,\hat{z})$ denotes the joint probability distribution of $z$ and $\hat{z}$.

The unknown conditional distribution $p(\hat{z}|z)$ can be approximated by the variational distribution $q_{\phi}(\hat{z}|z)$ with an auxiliary neural network parameterized with $\phi$~\cite{cheng2020club}. By feeding samples $\{(z_i, \hat{z}_i)\}_{i=1}^{N_\text{est}}$ into the auxiliary network, the MI between two feature distributions can be estimated as:
\begingroup
\small
\begin{align}
        \hat{I}(z ; \hat{z})=\frac{1}{{N_\text{est}}}\sum_{m=1}^{N_\text{est}} \left[\log q_\phi\left(\hat{z}_m|z_m\right)-\frac{1}{{N_\text{est}}}\sum_{n=1}^{N_\text{est}} \log q_\phi\left(\hat{z}_n|z_m\right)\right].
\end{align}
\endgroup

Previous works~\cite{guo2024domain,wang2021non} focused on minimizing the MI between the latent features output by the penultimate layer to reduce the generalization of $\mathcal{T}$. We observe that the source and target domain samples share minimal information at the penultimate layer. Their latent features in the earlier layers are fairly similar, which can leak vital information about the protected model's functionality. To avoid this vulnerability, we jointly minimize the MI between $\mathcal{S}$ and $\mathcal{T}$ on 4 selected hidden layers (i.e., $T_\text{sel} = \{l_1, l_2, l_3, l_4\}$) that are evenly distributed from shallow to deep within the model.

\section{The Proposed Method}

\begin{figure*}[t]
        \centering
        \includegraphics[width=1.0\linewidth]{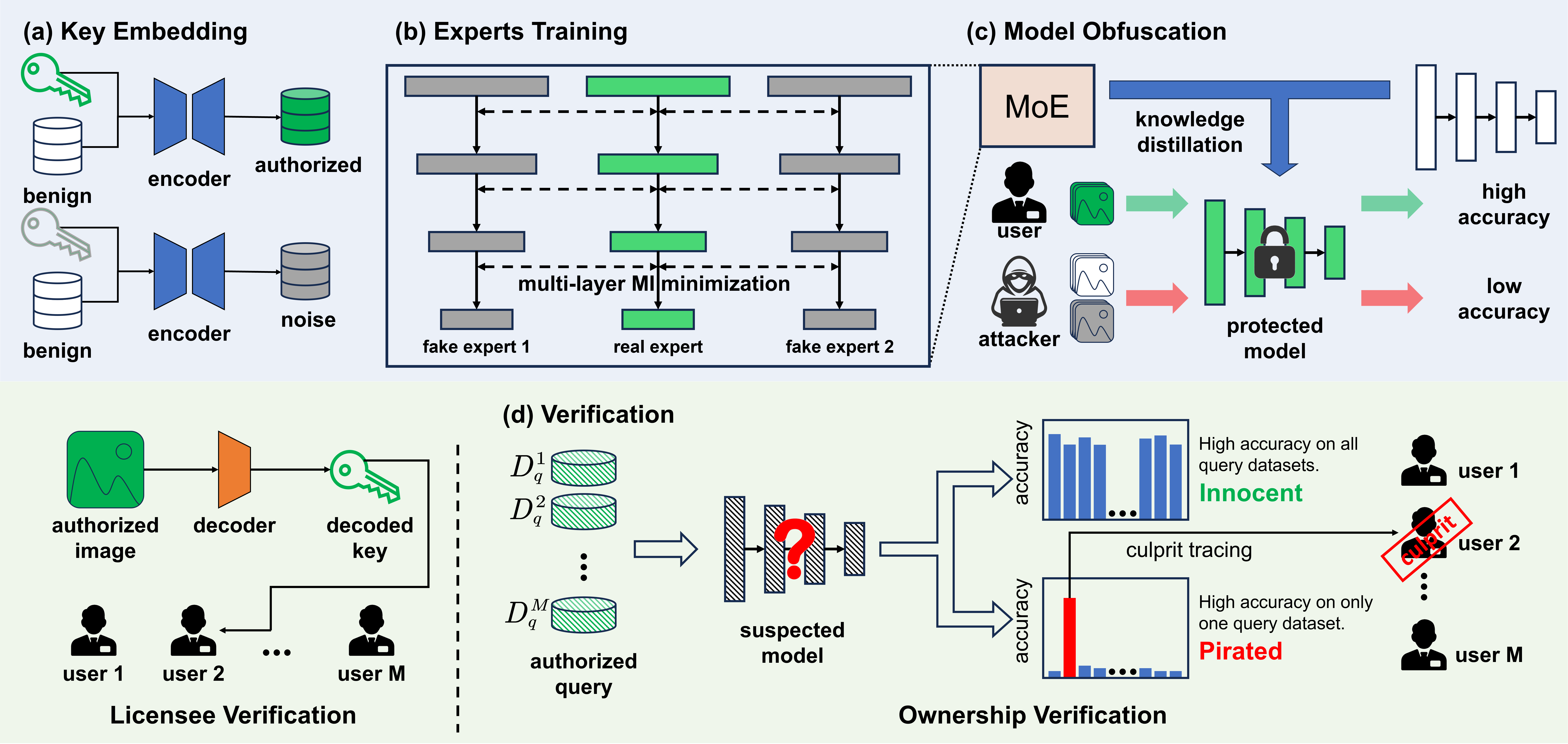}
        \caption{Overall pipeline of IDEA, a solution to the inverse problem of domain adaptation for key-based active authorization. The owner trains three experts: one real expert, which performs well on the authorized domain $\mathcal{A}$, and two fake experts, which behave like untrained models on the benign domain $\mathcal{B}$ and the noise domain $\mathcal{N}$, respectively. These three experts are seamlessly integrated into a unified model to enable active control while preventing the real expert's functionality from being leaked. To verify a licensee, the owner can extract a specific user key from encoded images provided by the user. Furthermore, the owner can conduct black-box queries on a suspected model to confirm the ownership and trace the dishonest user responsible for redeployment or infringement.}
        \label{fig:pipeline}
\end{figure*}

Fig.~\ref{fig:pipeline} illustrates the overall pipeline of our proposed method. It consists of three steps: key embedding, experts training, and model obfuscation. 

\subsection{Key Embedding} \label{sec:key_embedding}

\begin{figure}[t]
        \centering
        \includegraphics[width=1.0\columnwidth]{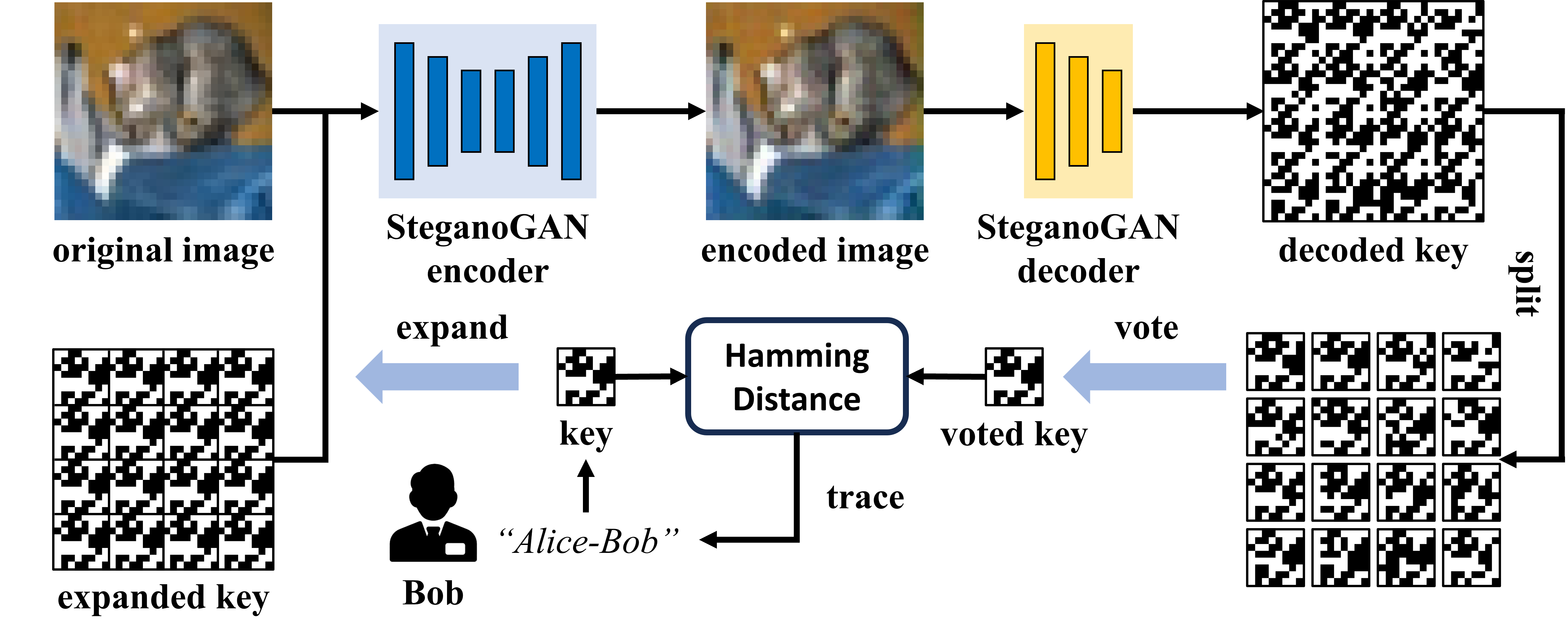}
        \caption{Key embedding into a cover image by the pretrained SteganoGAN encoder and key extraction from the stego image by the model owner with the corresponding pretrained SteganoGAN decoder (private). The SteganoGAN encoder and decoder are pretrained on the original training dataset.}
        \label{fig:encoding_decoding}
\end{figure}

Fig.~\ref{fig:encoding_decoding} illustrates the process of key encoding and decoding. Each user is assigned a unique key ${k}\in{\{0,1\}}^{{r}\times{r}}$, a binary square matrix where the first half encodes the owner's identity and the second half encodes the user's identity.

SteganoGAN requires the input image and the key to have matching dimensions in height and width. However, empirical observations indicate that directly matching the key size to the image size is inefficient. For high-resolution datasets like ImageNet, this approach results in an enormous key space of $2^{224\times224}$, which is unnecessary given the limited number of authorized users. Additionally, such a large key space hinders the protected model's ability to distinguish the correct key from incorrect ones, reducing the uniqueness of active authorization. To address this, the key size $r$ is fixed to values like 16 or 32, and multiple non-overlapping blocks of $k$ are concatenated to create an expanded key to match the image size. The encoded image is then generated by feeding the input image and the expanded key into the encoder simultaneously.

During the verification stage, the encoded image is passed through the decoder and split into multiple non-overlapping blocks of size ${r}\times{r}$. A majority voting process is then applied element-wise across the blocks to reconstruct the decoded key $\hat{k}$, which improves the decoding accuracy.

The encoder is provided to authorized users for embedding their unique keys into test images, while the decoder is kept private by the model owner to extract the embedded keys from encoded images. The perturbations introduced by the encoder are visually imperceptible and sample-specific, which implies that the perturbations vary across images even for the same embedded key.

\subsection{Experts Training} \label{sec:expert_training}

Let $\mathcal{A}$ be the source domain, and $\mathcal{B}$ and $\mathcal{N}$ be two different target domains. An inverse problem of domain adaption can be solved by optimizing the domain expected risk bound presented in Sec.~\ref{sec:domain_expected_risk_bound} to maintain the good performance of the authorized source domain and reduce the generalization of the unauthorized target domains. To fulfill \textbf{Definition 1}, the training objective of the protected model $f^\prime$ can be formulated as a bi-level optimization problem as follows:
\begingroup
\begin{equation}
    \begin{aligned}
        \mathop{\min} \quad & \mathbb{E}_{(x^+, y) \sim \mathcal{A}} \left[\mathcal{L}_{\text{CE}}\left(f^\prime(x^+), y\right)\right] \\
                            & + \lambda_1 \sum_{l \in T_\text{sel}} \left[\hat{I}\left(f^\prime_l(x^+); f^\prime_l(x)\right)\right] \\
                            & + \lambda_1 \sum_{l \in T_\text{sel}} \left[\hat{I}\left(f^\prime_l(x^+); f^\prime_l(x^-)\right)\right], \\
        \text{s.t.} \quad & \phi_{l,1} = \mathop{\arg\min}\limits_{\phi_{l,1}} \ {\hat{I}\left(f^\prime_l(x^+); f^\prime_l(x)\right)}, \\
                          & \phi_{l,2} = \mathop{\arg\min}\limits_{\phi_{l,2}} \ {\hat{I}\left(f^\prime_l(x^+); f^\prime_l(x^-)\right)},
        \label{eq:joint_optimization}
    \end{aligned}
\end{equation} 
\endgroup 
where $f^\prime_l(x)$, $f^\prime_l(x^+)$ and $f^\prime_l(x^-)$ denote the $l$-th layer latent features of $\mathcal{B}$, $\mathcal{A}$ and $\mathcal{N}$ domains, respectively. $\hat{I}(\cdot; \cdot)$ denotes the estimated MI between two domains. $\phi_{l,1}$ and $\phi_{l,2}$ are parameters of the two auxiliary networks used to estimate the $l$-th layer MI between the source domain and two target domains. $\lambda_1$ is a factor to balance the contributions of different terms. 

Unfortunately, it is computationally intensive to iteratively update the two auxiliary networks in tandem with the protected model. Moreover, the optimization of Eq.~\eqref{eq:joint_optimization} is difficult to converge. The trained model often exhibits polarized behavior, consistently classifying test images either correctly (satisfying the first term) or incorrectly (satisfying the last two terms), regardless of whether the images are authorized.

To address these issues, we design an MoE~\cite{jacobs1991adaptive} model $\mathcal{M}$ with three expert models: one real expert that performs well on $\mathcal{A}$, and two fake expert models that deliver random guessing performance on two unauthorized domains $\mathcal{B}$ and $\mathcal{N}$, respectively. The three experts have the same architecture as the unprotected model $f$.

The real expert $f_\text{real}$ is fine-tuned from $f$ for a few epochs to ensure that it achieves the optimal performance on $\mathcal{A}$ by optimizing
\begingroup
\begin{equation}
        \mathop{\min} \ {\mathbb{E}_{(x^+,y) \sim \mathcal{A}} \left[\mathcal{L}_\text{CE}\left(f_\text{real}(x^+),y\right)\right]}.
        \label{eq:real_expert}
\end{equation}
\endgroup

Conversely, the two fake experts, $f_\text{fake1}$ and $f_\text{fake2}$, are separately optimized by minimizing the MI of their latent representations with those of the real expert in selected layers as:
\begingroup
\begin{align}
        & \mathop{\min} \sum_{l \in T_\text{sel}} \hat{I}\left(f_{\text{real}, l}(x^+); f_{\text{fake1}, l}(x)\right), \label{eq:fake_expert_1} \\
        & \mathop{\min} \sum_{l \in T_\text{sel}} \hat{I}\left(f_{\text{real}, l}(x^+); f_{\text{fake2}, l}(x^-)\right), \label{eq:fake_expert_2}
\end{align}
\endgroup
where $f_{\text{real}, l}(x^+)$, $f_{\text{fake1}, l}(x)$, and $f_{\text{fake2}, l}(x^-)$ represent the $l$-th layer latent representations of $f_\text{real}$ on an authorized sample, $f_\text{fake1}$ on a benign sample, and $f_\text{fake2}$ on a noise sample, respectively.

The training objective of Eq.~\eqref{eq:joint_optimization} is decomposed into three terms independently optimized by Eqs.~\eqref{eq:real_expert},~\eqref{eq:fake_expert_1} and~\eqref{eq:fake_expert_2} in three separate models. Since the three optimization terms are now disentangled from each other, the training process of the three experts can converge more quickly, thereby reducing the training cost. The MoE model behaves as the real expert when fed with authorized images, but produces incorrect predictions, acting as one of the two fake experts when encountering benign or noise images.

\begin{figure}[t]
        \centering
        \includegraphics[width=0.95\columnwidth]{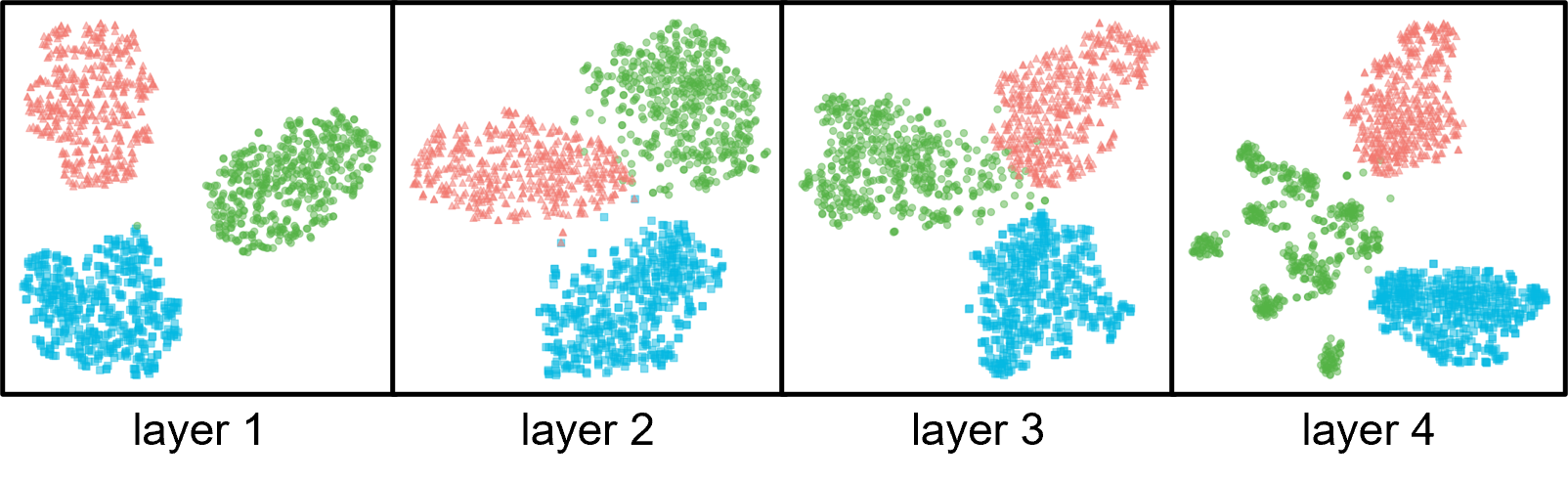}
        \caption{The t-SNE visualization~\cite{van2008visualizing} of latent outputs of the selected layers of MoE model to paired test images. The \textbf{\textcolor[HTML]{05b9e2}{blue}}, \textbf{\textcolor[HTML]{54b345}{green}}, and \textbf{\textcolor[HTML]{f27970}{red}} data points denote the latent representations of benign, authorized, and noise images, respectively.}
        \label{fig:MoE_representations}
\end{figure}

We separately feed 1500 paired test images sampled from $\mathcal{W} = \mathcal{A} \cup \mathcal{B} \cup \mathcal{N}$ (i.e., 500 samples for each subdomain) and visualize their latent features with t-distributed Stochastic Neighbor Embedding (t-SNE) technique~\cite{van2008visualizing}. As shown in Fig.~\ref{fig:MoE_representations}, the latent features form three distinct clusters corresponding to their respective subdomains. As expected, the authorized domain shares minimal MI with the two unauthorized domains in the latent space. Consequently, the latent features from different domains exhibit increased separation.

\begin{algorithm}[t]
        \caption{Encoded Model Generation}
        \label{alg:generation}
        \LinesNumbered
        
        \KwIn{
            Benign training dataset $D_\text{tr}$;
            unprotected model $f$;
            SteganoGAN encoder $\mathcal{E}$;
            $M$ unique user keys $\mathcal{K}=\{k_1, k_2,\cdots,k_M\}$;
            training epochs $\{E_1, E_2, E_3\}$;
            learning rates $\{\eta_1, \eta_2, \eta_3\}$;
            set of selected hidden layers $T_\text{sel}$.
        }      
        
        \KwOut{$M$ model-key pairs.}
         
        \For{$k \in \{k_1, k_2,\cdots,k_M\}$}{

            \textcolor{blue}{\# Generate the authorized and noise training datasets} \\
            $D^+_{\text{tr}, j} \gets \{(\mathcal{E}(x_i, k), y_i)\}^{N_\text{tr}}_{i=1}$ \\
            $D^-_{\text{tr}, j} \gets \{(\mathcal{E}(x_i, k^{\ast}), y_i)\}^{N_\text{tr}}_{i=1}$ \\ 
            $\overline{D}_{\text{tr}, j} = D_\text{tr} + D^+_{\text{tr}, j} + D^-_{\text{tr}, j}$

            \textcolor{blue}{\# Experts training} \\
            Fine-tune the real expert model for $E_1$ epochs from the original unprotected model. \\
            \Comment{Eq.~\eqref{eq:real_expert}} \\

            Separately optimize two randomly initialized models for $E_2$ epochs to obtain the fake experts. \\
            \Comment{Eq.~\eqref{eq:fake_expert_1}~\eqref{eq:fake_expert_2}} \\

            Construct the MoE model $\mathcal{M}$.\\
            
            \textcolor{blue}{\# Model Obfuscation} \\
            Distill the MoE teacher model for $E_3$ epochs to generate the protected student model $f^\prime$. \\
            \Comment{Eq.~\eqref{eq:obf_loss}} \\
                
            \Return $f^\prime$.
            }
        Distribute each model-key pair and the encoder (i.e., $\{f^\prime, k\}$ and $\mathcal{E}$) to the corresponding legal user.
\end{algorithm}

\subsection{Model Obfuscation} \label{sec:model_obfuscation}

It is worth noting that the MoE model $\mathcal{M}$ lacks a gating network to decide which expert should be activated for prediction. Consequently, $\mathcal{M}$ requires prior knowledge about the domain from which the input image is sampled to activate the correct domain expert, which is impractical in real-world scenarios. Additionally, distributing $\mathcal{M}$ directly to a user poses a security risk, as the adversary can easily identify the real expert due to its superior classification performance on benign test samples compared to the two fake experts. To address these issues, we obfuscate the three experts by unifying them into a single coherent model using knowledge distillation. We use $\mathcal{M}$ to train a student model $f^\prime$ which imitates the teacher's behaviors on not only the fully-connected layer but all selected hidden layers from shallow to deep. To achieve this, $f^\prime$ is jointly optimized with three knowledge distillation losses $\mathcal{L}_\text{kl}$, $\mathcal{L}_\text{at}$ and $\mathcal{L}_\text{crd}$ such that
\begingroup
\begin{equation}
        \mathop{\min} \left[ \mathcal{L}_\text{kl} + \lambda_2 \mathcal{L}_\text{at} + \lambda_3 \mathcal{L}_\text{crd} \right],
        \label{eq:obf_loss}
\end{equation}
\endgroup
where $\lambda_2$ and $\lambda_3$ are two positive factors to adjust the scale of $\mathcal{L}_\text{at}$ and $\mathcal{L}_\text{crd}$, respectively.

(1) Minimizing $\mathcal{L}_\text{kl}$ aligns the logits output by the fully connected layers of the teacher and student models. $\mathcal{L}_\text{kl}$ can be computed as:
\begingroup
\begin{equation}
        \mathcal{L}_\text{kl} = \mathbb{E}_{x \sim \mathcal{W}} \left[\mathcal{L}_\text{KL}\left(f^\prime(x), \mathcal{M}(x)\right)\right]
        \label{eq:kl_loss}
\end{equation}
\endgroup
where $\mathcal{L}_\text{KL}$ denotes the Kullback-Leibler (KL) divergence between two distributions.

(2) A 3D feature map $f_l \in \mathbb{R}^{{c}\times{h}\times{w}}$ can be flattened along the channel dimension as $\mathcal{A}^\alpha(f_l) = \sum^{c}_{i=1}\left|f_{l,(i)}\right|^\alpha$, where $\alpha > 1$ is the factor that amplifies large neuron activations. Enlarging $\alpha$ places greater emphasis on regions with high output values. $\alpha$ is fixed to 2 according to~\cite{li2021neural}. 

The attention transfer loss~\cite{zagoruyko2016paying} is computed as:
\begingroup
\begin{equation}
        \begin{aligned}
                \mathcal{L}_\text{AT}(f_l^\text{T}, f_l^\text{S}) = \left\|\frac{\mathcal{A}^\alpha\left(f_l^\text{T}\right)}{\left\|\mathcal{A}^\alpha\left(f_l^\text{T}\right)\right\|_2}-\frac{\mathcal{A}^\alpha\left(f_l^\text{S}\right)}{\left\|\mathcal{A}^\alpha\left(f_l^\text{S}\right)\right\|_2}\right\|_2,
                \label{eq:attention_transfer}
        \end{aligned}
\end{equation}
\endgroup
where $f_l^\text{T}$ and $f_l^\text{S}$ represent the features extracted from the teacher and student models, respectively at the same layer $l$. We also align the neuron attention of teacher $\mathcal{M}$ and the student $f^\prime$ on the selected layers by minimizing 
\begingroup
\begin{equation}
        \mathcal{L}_\text{at} = \mathbb{E}_{x \sim \mathcal{W}} \left[ \sum_{l \in T_\text{sel}} \mathcal{L}_\text{AT}\left(\mathcal{M}_l(x), f^\prime_l(x)\right)\right].
        \label{eq:at_loss}
\end{equation}
\endgroup

(3) To ensure that input samples from different domains also share low MI in the latent space in the student model, contrastive representation distillation (CRD)~\cite{tian2019contrastive} loss is incorporated. Given a tuple of teacher and student representations $(f_l^T(x_a),f_l^S(x_b))$, where $x_a$ and $x_b$ are two random images sampled from $\mathcal{W}$. It is a \textit{positive} pair if the two inputs are from the same domain (e.g., $x_a \sim \mathcal{A}$ and $x_b \sim \mathcal{A}$). Conversely, it is a \textit{negative} pair if the two inputs are sourced from different domains (e.g., $x_a \sim \mathcal{A}$ and $x_b \sim \mathcal{B}$). Let $q_1$ and $q_0$ denote the joint distributions for \textit{positive} and \textit{negative} pairs, respectively. The CRD loss is given by:
\begingroup
\begin{align}
        \mathcal{L}_\text{crd} = & - \sum_{l \in T_\text{sel}} \mathbb{E}_{q_1} \left[\text{log}\left(h_l(f_l^T,f_l^S)\right)\right] \\
                                 & - N_\text{neg} \sum_{l \in T_\text{sel}} \mathbb{E}_{q_0} \left[\text{log}\left(1-h_l(f_l^T,f_l^S)\right)\right],
        \label{eq:crd_loss}
\end{align}
\endgroup
where $h_l(\cdot,\cdot)$ denotes a critic to determine if the two representations are positively paired, and whose parameters can be optimized in parallel with $\mathcal{L}_\text{crd}$. We provide one \textit{positive} pair for every $N_\text{neg}$ \textit{negative} pairs during training, with the aim to pull the \textit{positive} pairs closer and push the \textit{negative} pairs apart.

The protected model $f^\prime$ has similar performance as the MoE model $\mathcal{M}$, but the functionalities and hidden-layer behaviors of the three experts have been \textbf{indistinguishably} blent into a coherent network, making it infeasible for the adversary to extract the real expert. The algorithm for generating $M$ model-key pairs is depicted in Algorithm~\ref{alg:generation}.

\subsection{Verification} \label{sec:verification_tracing}

\textbf{Licensee Verification.} To securely demonstrate a legitimate licensee without revealing sensitive information, an authorized user provides one or more encoded images instead of directly sharing the secret key. The SteganoGAN decoder, which is kept private by the owner, is used to extract the embedded key from these images. The licensee is validated if the Hamming distance (HD) between the extracted key and a recorded key in the owner's database falls below a predefined threshold.

Let $D_\text{e}$ be an encoded dataset where each image is encoded with a random key, the decoding accuracy (DA) is defined as:
\begingroup
\begin{equation}
        \text{DA} = \mathbb{E}_{x \sim D_\text{e}} \left[\mathbb{I}\{\text{HD}\left(k, \hat{k}\right) \leq \epsilon_3\}\right],
        \label{eq:tsr}
\end{equation}
\endgroup
where $\epsilon_3$ is the predefined threshold, $k$ and $\hat{k}$ represent the actual and extracted keys, respectively. A high DA indicates that the decoder can accurately extract embedded keys from encoded images to provide an accurate and secure licensee verification.

\textbf{Ownership Verification.} The owner can conduct a series of black-box queries and analyze the outputs to determine whether $\hat{f}$ is an illegally redeployed model. If unauthorized redistribution is confirmed, the dishonest user responsible can be identified.

\textbf{Definition 2} (Black-box Query-based Ownership Verification). \textit{Let $D^\mathcal{B}_q$ denote a benign query set sampled from the benign domain. Ownership over the suspected model $\hat{f}$ is verified if and only if there exists exactly one $k \in \mathcal{K}$ satisfying:
\begingroup
\begin{align}
      & \left| \operatorname{Acc}\{f, D^\mathcal{B}_q\} - \operatorname{Acc}\{\hat{f}, D^\mathcal{A}_q\} \right| < \epsilon_{1}, \label{eq:4}\\
      & \operatorname{Acc}\{f, D^\mathcal{B}_q\} - \operatorname{Acc}\{\hat{f}, D^\mathcal{N}_q\} > \epsilon_{2}, \label{eq:5}
\end{align}
\endgroup
where $D^\mathcal{A}_q$ consists of query samples embedded with $k$, and $D^\mathcal{N}_q$ consists of query samples embedded with other recorded keys.}

Eq.~\eqref{eq:4} proves that the full functionality of the suspected model is activated by a specific recorded key, and Eq.~\eqref{eq:5} confirms that $\hat{f}$ cannot be an innocent model due to its degraded performance on non-matching keys. Ownership is thereby validated, and the dishonest user is identified, as the traced key encodes the identities of both the owner and the user.

\begin{table}[t]
        \centering
        \renewcommand\arraystretch{1.0}
        \caption{Specifications of the experimental datasets.}
        \vspace{-2mm}
        \resizebox{0.95\columnwidth}{!}{
          \begin{tabular}{@{\hspace{0.2cm}}cccc@{\hspace{0.2cm}}}
            \toprule
            Dataset        & \# Labels & Input Size       & \# Training / Test Images       \\ \midrule
            CIFAR-10       & 10        & 32 $\times$ 32   & 50000       / 10000             \\
            GTSRB          & 43        & 32 $\times$ 32   & 39208       / 12630             \\
            Caltech-101    & 101       & 32 $\times$ 32   & 6942        / 1735              \\
            ImageNet-10    & 10        & 224 $\times$ 224 & 12000       / 1000              \\ \bottomrule
          \end{tabular}
        }
        \label{tab:datasets}
\end{table}

\section{Experiments} \label{sec:experiments}

\subsection{Experimental Setting}

\subsubsection{Datasets and Networks}

Table~\ref{tab:datasets} specifies the four image classification datasets used in our experiments: CIFAR-10~\cite{krizhevsky2009learning}, Caltech-101~\cite{FeiFei2004LearningGV}, GTSRB~\cite{houben2013detection}, and ImageNet~\cite{deng2009imagenet}, with different resolutions and numbers of classes. To speed up the training without loss of generality, ImageNet evaluation is performed on a subset of 10 randomly selected classes, referred to as ImageNet-10. Four popular architectures are employed in our evaluation: ResNet-18 (Res-18)~\cite{he2016deep}, PreActResNet-18 (PreRes-18)~\cite{he2016deep}, SENet~\cite{hu2018squeeze} and MobileNet-V2 (Mob-V2)~\cite{howard2017mobilenets}.

\subsubsection{Other settings}

The key size parameter $r$ is set to 32 for ImageNet-10, and 16 for the other four datasets. For each classification task, a dedicated SteganoGAN model is trained using 2500 samples randomly selected from the benign training dataset. The unprotected model $f$ is trained from scratch for 200 epochs, and from which the real expert $f_\text{real}$ is fine-tuned for 40 epochs, both using the SGD optimizer. Each fake expert is separately trained for 1000 iterations. During the knowledge distillation process, the student model is optimized from scratch for 50 epochs using the Adam optimizer. Ten protected models are generated from the same source model with different random keys for distribution to ten different users. Several common data augmentations, including random cropping, random horizontal flipping, random rotation and random erasing, are applied during training to alleviate overfitting.

\begin{figure*}[t]
        \centering
        \includegraphics[width=1.0\linewidth]{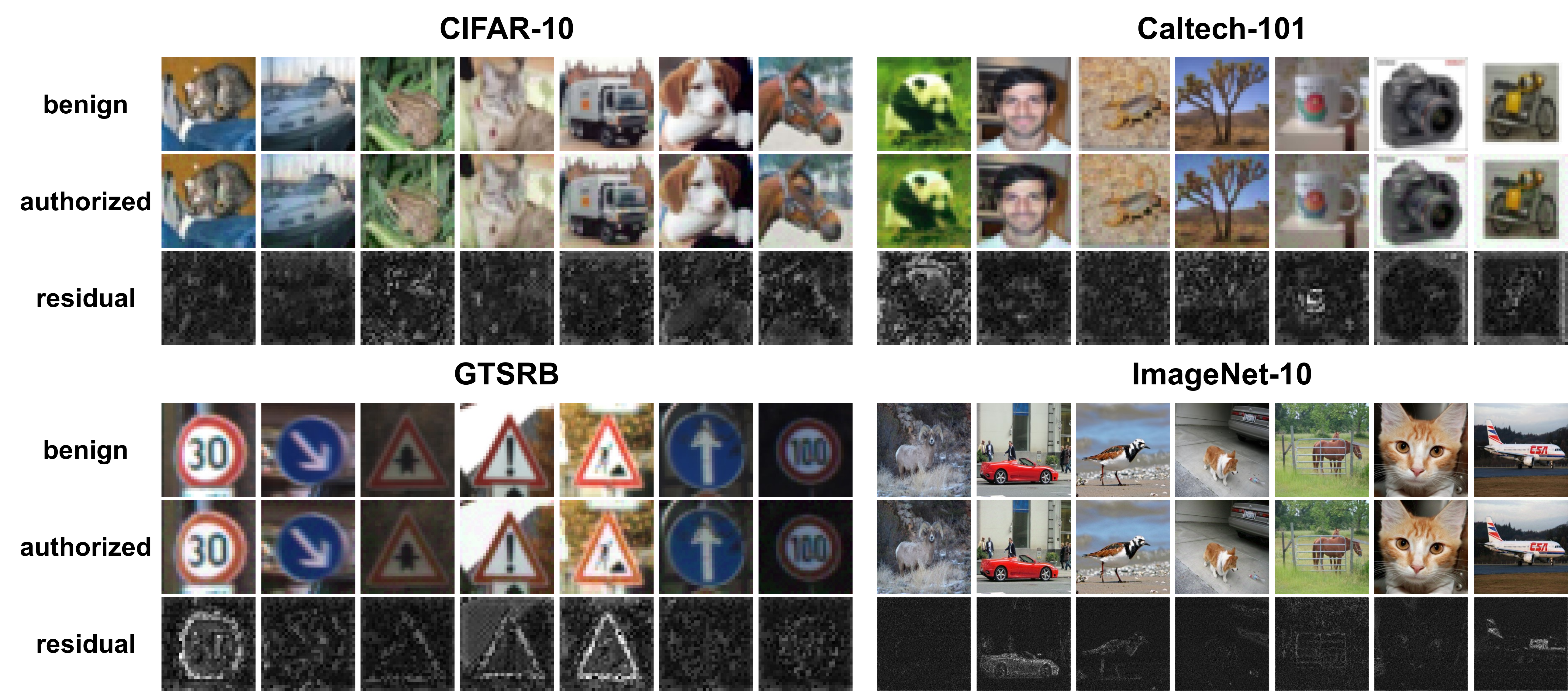}
        \caption{Examples of benign (first row) and authorized (second row) images. For each dataset, all authorized images are embedded with the same binary key, and the magnified ($\times 10$) residual images are shown in the third row.}
        \label{fig:visualization}
\end{figure*}

\begin{table}[h]
        \caption{IQA scores of authorized images. $\uparrow$ denotes larger value is better, and $\downarrow$ denotes smaller value is better.}
        \centering
        \renewcommand\arraystretch{1.0}
        \resizebox{0.95\linewidth}{!}{
            \begin{tabular}{@{\hspace{0.2cm}}cccc@{\hspace{0.2cm}}}
                \toprule
                \multirow{2}{*}{Dataset} & \multicolumn{3}{c}{IQA Metric}                                   \\ \cmidrule(l){2-4} 
                                         & SSIM ($\uparrow$)   & PSNR ($\uparrow$)    & LPIPS ($\downarrow$)\\ \midrule
                CIFAR-10                 & 0.9928 $\pm$ 0.0050 & 33.1038 $\pm$ 1.5923 & 0.0032 $\pm$ 0.0025 \\
                GTSRB                    & 0.9914 $\pm$ 0.0129 & 32.4346 $\pm$ 2.1956 & 0.0032 $\pm$ 0.0028 \\
                Caltech-101              & 0.9882 $\pm$ 0.0082 & 33.9264 $\pm$ 2.2390 & 0.0053 $\pm$ 0.0030 \\
                ImageNet-10              & 0.9706 $\pm$ 0.0249 & 35.3929 $\pm$ 1.4052 & 0.0144 $\pm$ 0.0089 \\ \bottomrule
            \end{tabular}
        }
        \label{tab:IQA_score}
\end{table}

\subsection{Evaluation} \label{sec:evaluation}

\subsubsection{Stealth}

Fig.~\ref{fig:visualization} presents some examples of benign and authorized images for CIFAR-10, Caltech-101, GTSRB, and ImageNet-10 datasets. We also provide the magnified ($\times10$) residual images obtained by the absolute difference in pixel values between the benign and authorized images. The perturbations introduced by SteganoGAN encoding are visually imperceivable. In other words, the attacker can hardly figure out the embedded key by inspecting a few test images encoded with the same key. Furthermore, the perturbations are sample-specific. Even if the attacker can compute a residual image from a pair of benign and authorized images of a protected model, it is impossible to generate an authorized image by adding the residual image to a different test image. The means and standard deviations of three objective image quality assessments (IQA), namely Structural Similarity Index Measure (SSIM)~\cite{hore2010image}, Peak Signal to Noise Ratio (PSNR) and Learned Perceptual Image Patch Similarity (LPIPS)~\cite{zhang2018unreasonable}, are presented in Table~\ref{tab:IQA_score}. The outstanding IQA scores prove the stealth of the key-encoded images.

\begin{table*}[t]
        \caption{Prediction accuracy (\%) of unprotected and protected models on benign/authorized/noise test images across different networks and datasets. For each case, the unprotected model's performance on benign dataset is highlighted in \textbf{\textcolor[HTML]{05b9e2}{blue}} and set as the baseline; the protected model's performance on authorized dataset is highlighted in \textbf{\textcolor[HTML]{54b345}{green}}. The acc. change (\%) is calculated by subtracting the former from the latter.}
        \centering
        \renewcommand\arraystretch{1.4}
        \resizebox{1.0\linewidth}{!}{
                \begin{tabular}{c|c|cccc}
                        \Xhline{2\arrayrulewidth}
                        \multirow{2}{*}{Dataset}     & \multirow{2}{*}{Model} & \multicolumn{4}{c}{Test on benign / authorized / noise images(\%)} \\ \cline{3-6} 
                                                &                   & \multicolumn{1}{c|}{Res-18}                & \multicolumn{1}{c|}{PreRes-18}             & \multicolumn{1}{c|}{SENet}                 & Mob-V2                \\ \Xhline{2\arrayrulewidth}
                        \multirow{3}{*}{CIFAR-10}    & unprotected       & \multicolumn{1}{c|}{\textbf{\textcolor[HTML]{05b9e2}{93.58}} / 91.87 / 91.99} & \multicolumn{1}{c|}{\textbf{\textcolor[HTML]{05b9e2}{93.82}} / 91.83 / 92.08} & \multicolumn{1}{c|}{\textbf{\textcolor[HTML]{05b9e2}{93.68}} / 92.07/ 92.15}  & \textbf{\textcolor[HTML]{05b9e2}{91.63}} / 89.66 / 89.84 \\ 
                                                & protected         & \multicolumn{1}{c|}{10.13 / \textbf{\textcolor[HTML]{54b345}{93.15}} / 9.69}  & \multicolumn{1}{c|}{12.66 / \textbf{\textcolor[HTML]{54b345}{93.00}} / 14.70} & \multicolumn{1}{c|}{1.79 / \textbf{\textcolor[HTML]{54b345}{93.10}} / 12.18}  & 10.06 / \textbf{\textcolor[HTML]{54b345}{91.37}} / 12.72 \\ \cline{2-6} 
                                                & acc. change (\%)     & \multicolumn{1}{c|}{-0.43}                 & \multicolumn{1}{c|}{-0.82}                 & \multicolumn{1}{c|}{-0.58}                 & -0.26                 \\ \hline
                        \multirow{3}{*}{Caltech-101} & unprotected       & \multicolumn{1}{c|}{\textbf{\textcolor[HTML]{05b9e2}{75.73}} / 75.73 / 76.02} & \multicolumn{1}{c|}{\textbf{\textcolor[HTML]{05b9e2}{76.71}} / 76.60 / 76.37} & \multicolumn{1}{c|}{\textbf{\textcolor[HTML]{05b9e2}{76.08}} / 75.68 / 75.79} & \textbf{\textcolor[HTML]{05b9e2}{75.91}} / 75.22 / 75.79 \\  
                                                & protected         & \multicolumn{1}{c|}{1.04 / \textbf{\textcolor[HTML]{54b345}{77.23}} / 1.38}   & \multicolumn{1}{c|}{1.56 / \textbf{\textcolor[HTML]{54b345}{77.81}} / 1.27}   & \multicolumn{1}{c|}{0.40 / \textbf{\textcolor[HTML]{54b345}{75.62}} / 1.10}   & 1.90 / \textbf{\textcolor[HTML]{54b345}{76.02}} / 2.59   \\ \cline{2-6} 
                                                & acc. change (\%)     & \multicolumn{1}{c|}{1.50}                  & \multicolumn{1}{c|}{1.10}                  & \multicolumn{1}{c|}{-0.46}                 & 0.11                  \\ \hline
                        \multirow{3}{*}{GTSRB}       & unprotected       & \multicolumn{1}{c|}{\textbf{\textcolor[HTML]{05b9e2}{97.29}} / 92.26 / 92.98} & \multicolumn{1}{c|}{\textbf{\textcolor[HTML]{05b9e2}{98.01}} / 92.77 / 92.89} & \multicolumn{1}{c|}{\textbf{\textcolor[HTML]{05b9e2}{97.11}} / 91.94 / 93.86} & \textbf{\textcolor[HTML]{05b9e2}{98.74}} / 97.24 / 97.14 \\ 
                                                & protected         & \multicolumn{1}{c|}{2.01 / \textbf{\textcolor[HTML]{54b345}{97.78}} / 3.33}   & \multicolumn{1}{c|}{5.34 / \textbf{\textcolor[HTML]{54b345}{97.95}} / 5.57}   & \multicolumn{1}{c|}{1.15 / \textbf{\textcolor[HTML]{54b345}{97.75}} / 2.68}   & 6.94 / \textbf{\textcolor[HTML]{54b345}{98.12}} / 5.84   \\ \cline{2-6} 
                                                & acc. change (\%)     & \multicolumn{1}{c|}{0.49}                  & \multicolumn{1}{c|}{-0.06}                 & \multicolumn{1}{c|}{0.64}                  & -0.62                 \\ \hline
                        \multirow{3}{*}{ImageNet-10} & unprotected       & \multicolumn{1}{c|}{\textbf{\textcolor[HTML]{05b9e2}{88.85}} / 88.69 / 88.81} & \multicolumn{1}{c|}{\textbf{\textcolor[HTML]{05b9e2}{89.38}} / 89.23 / 89.38} & \multicolumn{1}{c|}{\textbf{\textcolor[HTML]{05b9e2}{88.81}} / 88.58 / 88.77} & \textbf{\textcolor[HTML]{05b9e2}{85.92}} / 86.27 / 85.96 \\ 
                                                & protected         & \multicolumn{1}{c|}{10.35 / \textbf{\textcolor[HTML]{54b345}{89.31}} / 10.08} & \multicolumn{1}{c|}{10.46 / \textbf{\textcolor[HTML]{54b345}{87.88}} / 11.96} & \multicolumn{1}{c|}{9.88 / \textbf{\textcolor[HTML]{54b345}{89.42}} / 9.04}   & 10.31 / \textbf{\textcolor[HTML]{54b345}{87.69}} / 11.31 \\ \cline{2-6} 
                                                & acc. change (\%)     & \multicolumn{1}{c|}{0.46}                  & \multicolumn{1}{c|}{-1.50}                 & \multicolumn{1}{c|}{0.61}                  & 1.77                  \\ \Xhline{2\arrayrulewidth}
                \end{tabular}
        }
        \label{tab:accuracy}
        \vspace{-4mm}
\end{table*}

\subsubsection{Effectiveness and fidelity}

The inference performances of both unprotected and protected models on benign, authorized, and noise test images are presented in Table~\ref{tab:accuracy}. The results indicate that unprotected models consistently achieve high prediction accuracy across all three domains. All protected models fulfill the \textbf{fidelity} requirement of Eq.~\eqref{eq:fidelity}, with inference accuracy of authorized images comparable to the benign accuracy of the corresponding unprotected models. The accuracy drop is no more than 1.5\%. Additionally, the protected models perform poorly on benign and noise images, with near-random guessing accuracies, making them useless to unauthorized users without the correct key.

\begin{figure}[t]
        \centering
        \begin{subfigure}{0.48\columnwidth}
                \includegraphics[width=\linewidth]{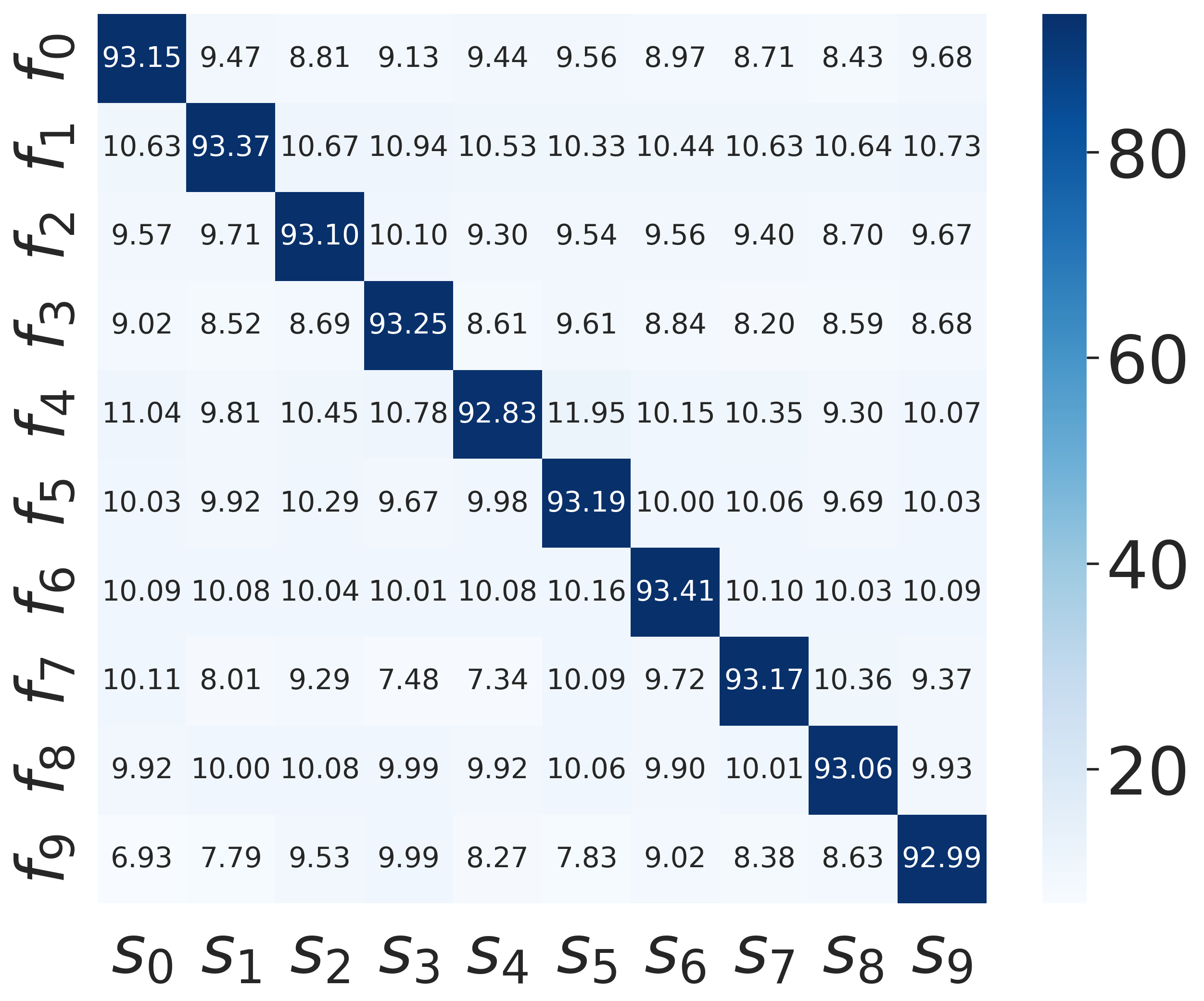}
                \caption{CIFAR-10}
        \end{subfigure}
        \begin{subfigure}{0.48\columnwidth}
                \includegraphics[width=\linewidth]{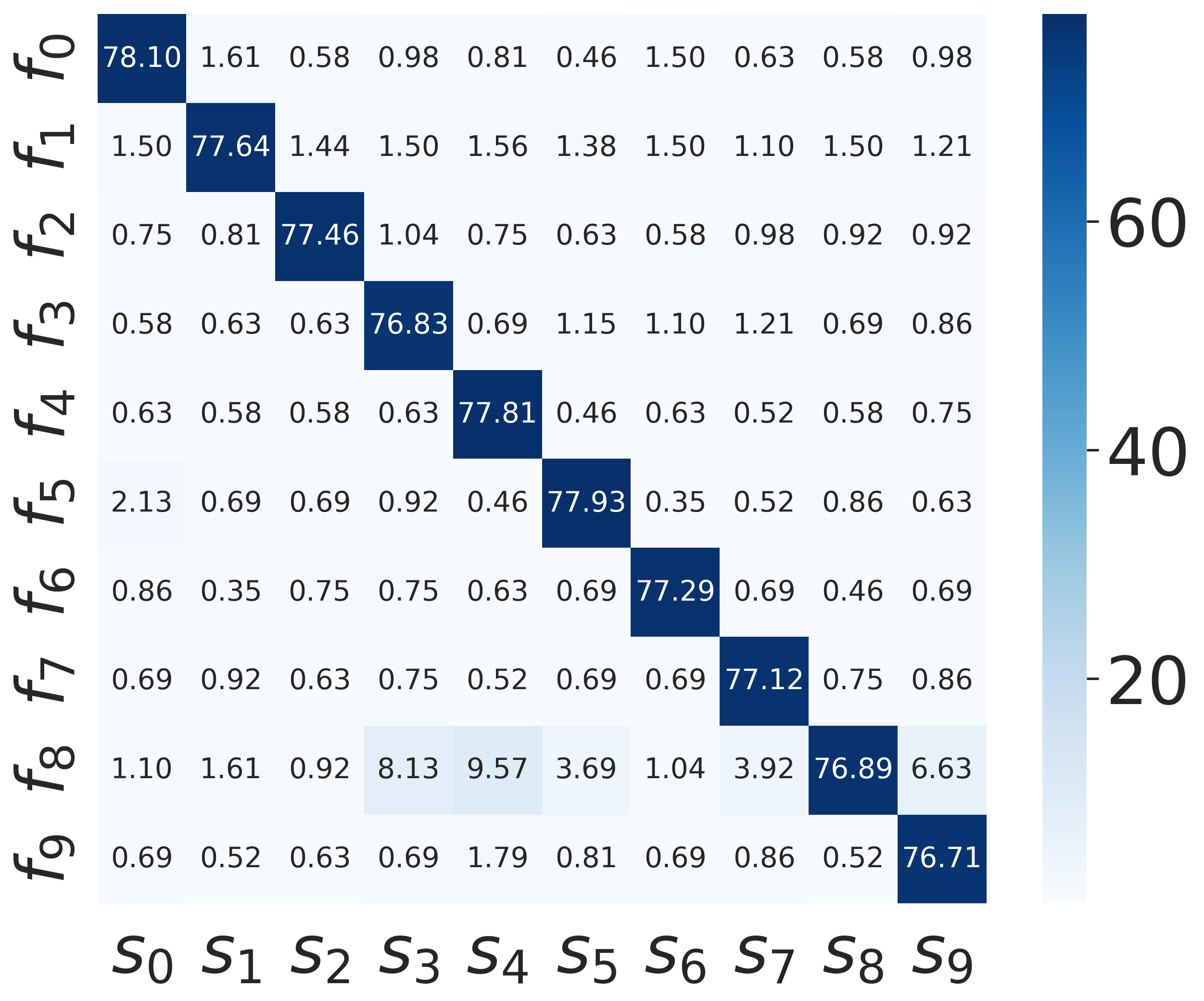}
                \caption{Caltech-101}
        \end{subfigure}
        \\
        \begin{subfigure}{0.48\columnwidth}
                \includegraphics[width=\columnwidth]{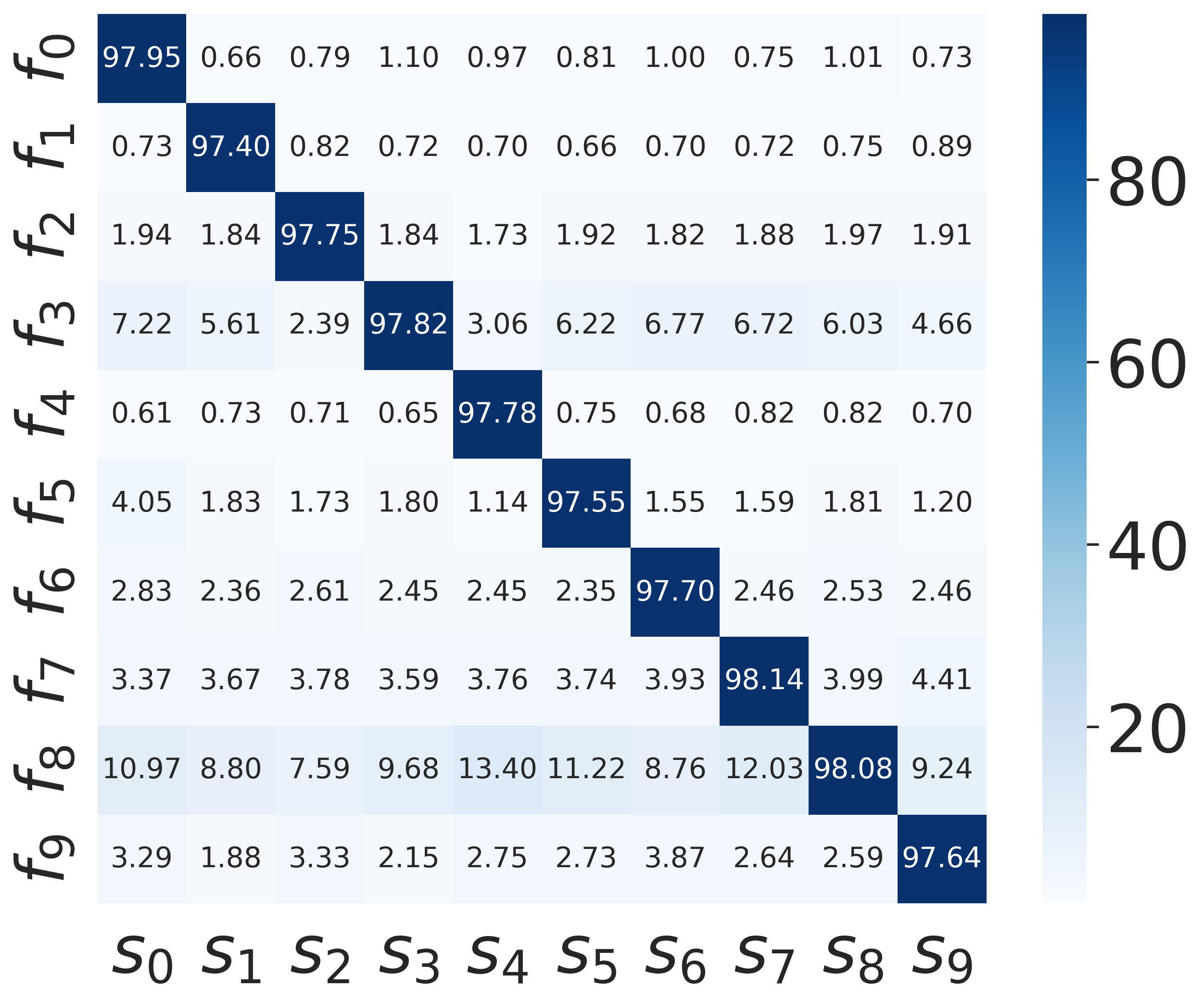}
                \caption{GTSRB}
        \end{subfigure}
        \begin{subfigure}{0.48\columnwidth}
                \includegraphics[width=\columnwidth]{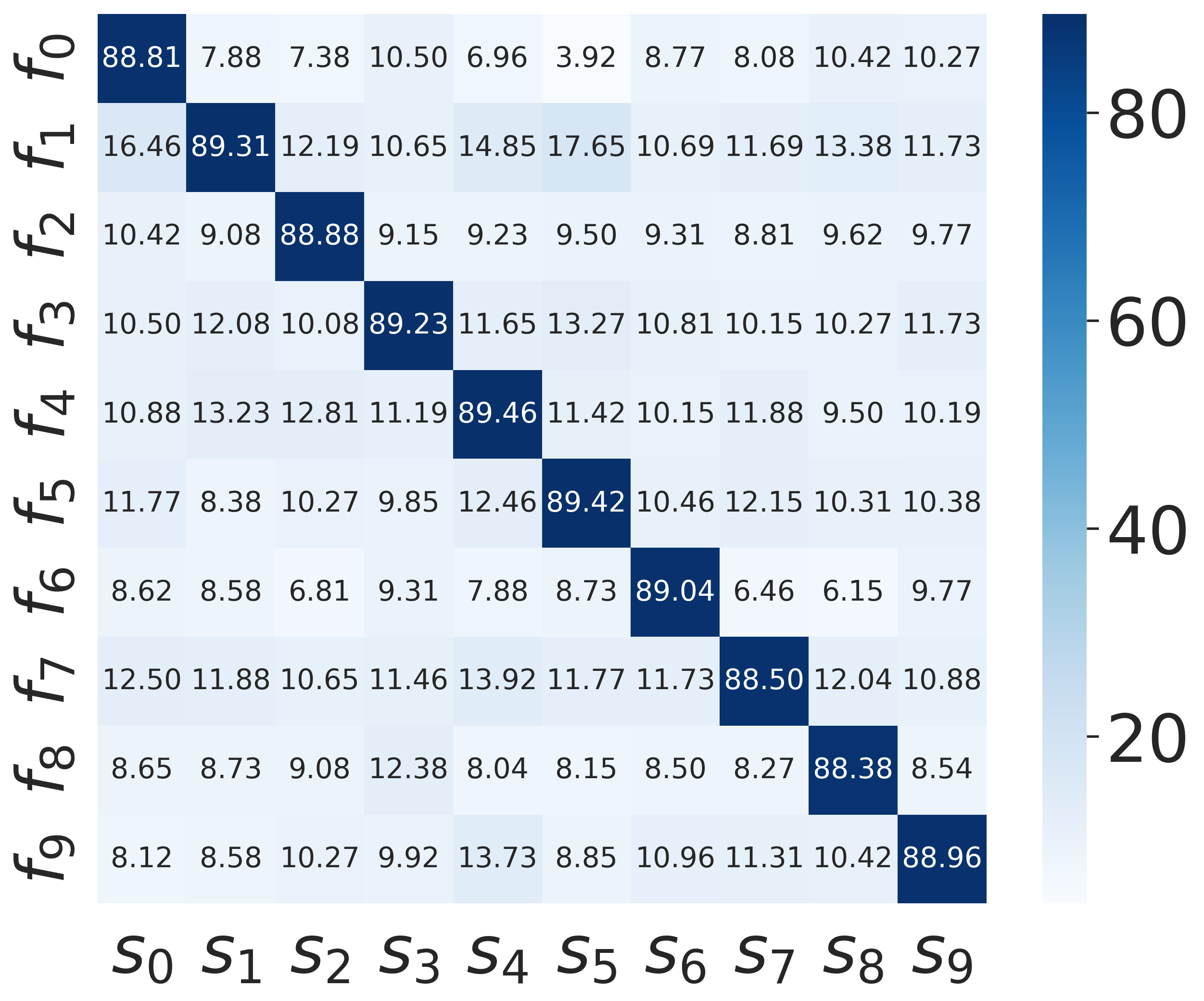}
                \caption{ImageNet-10}
        \end{subfigure}

        \caption{Confusion matrices for the evaluation of 10 different keys on the 10 encoded models. The model is ResNet-18.}

        \label{fig:confusion_matrices}
\end{figure}

\begin{figure}[t]
        \centering
        \begin{subfigure}{0.48\columnwidth}
                \includegraphics[width=\columnwidth]{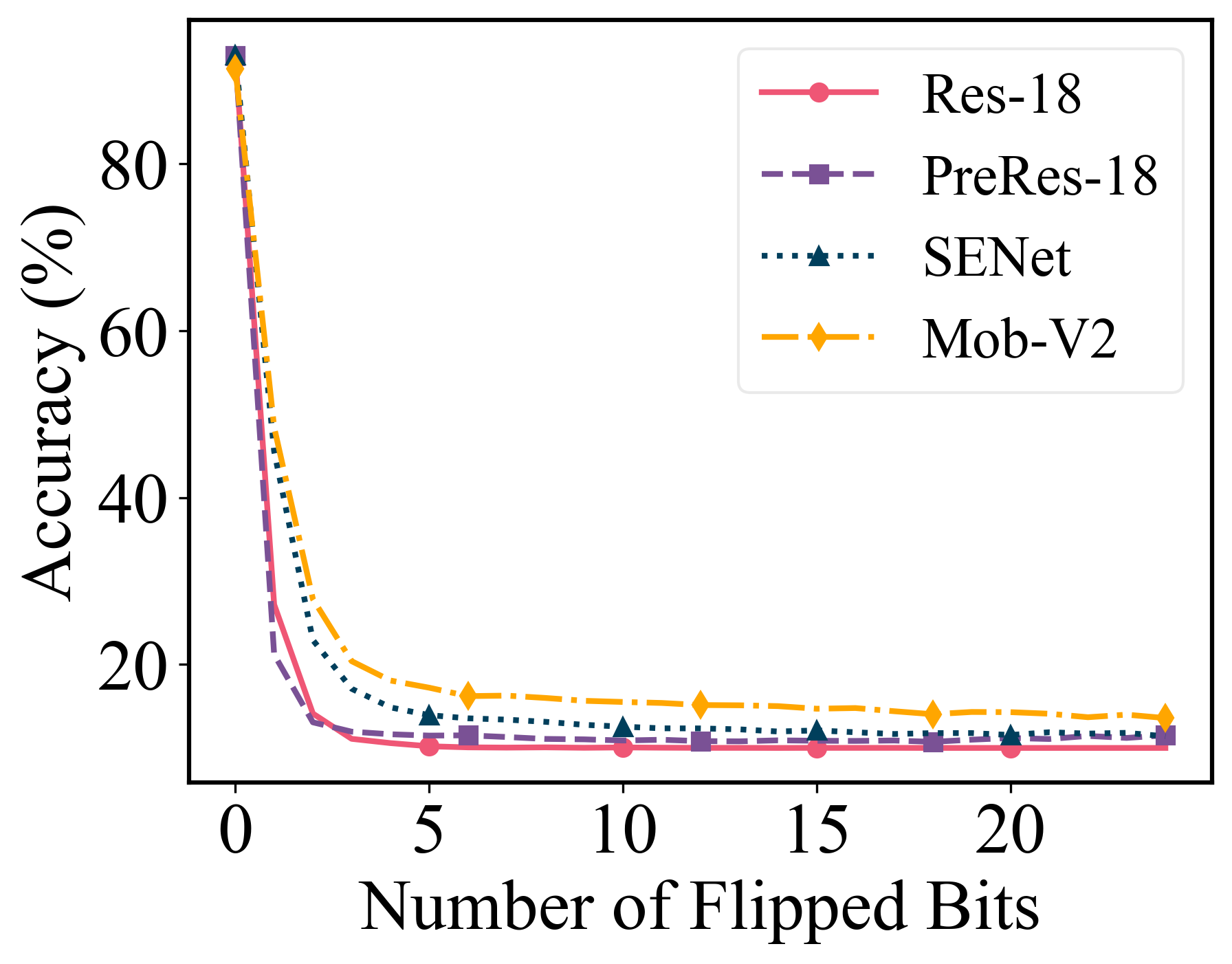}
                \caption{CIFAR-10}
        \end{subfigure}
        \begin{subfigure}{0.48\columnwidth}
                \includegraphics[width=\columnwidth]{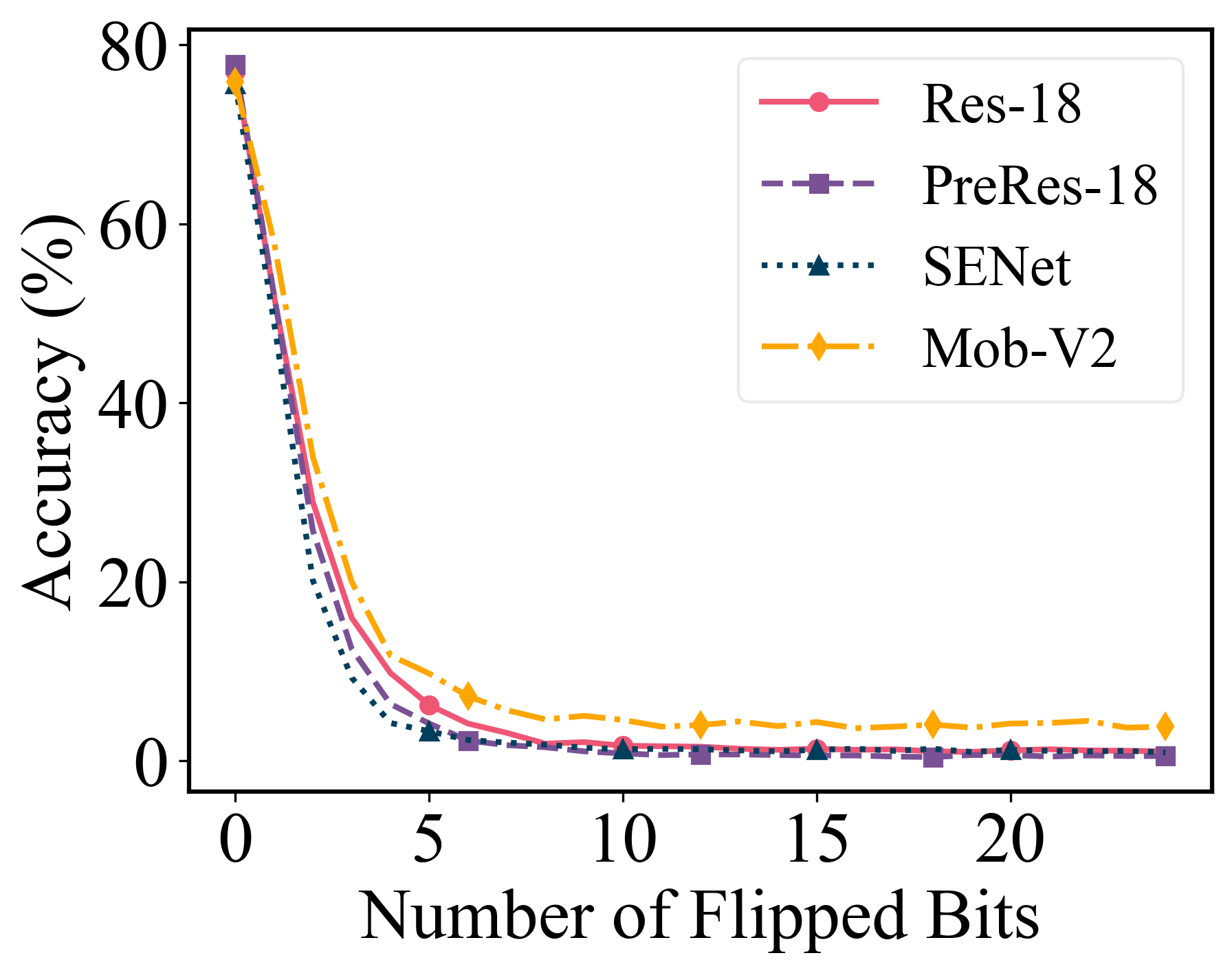}
                \caption{Caltech-101}
        \end{subfigure}
        \\
        \begin{subfigure}{0.48\columnwidth}
                \includegraphics[width=\columnwidth]{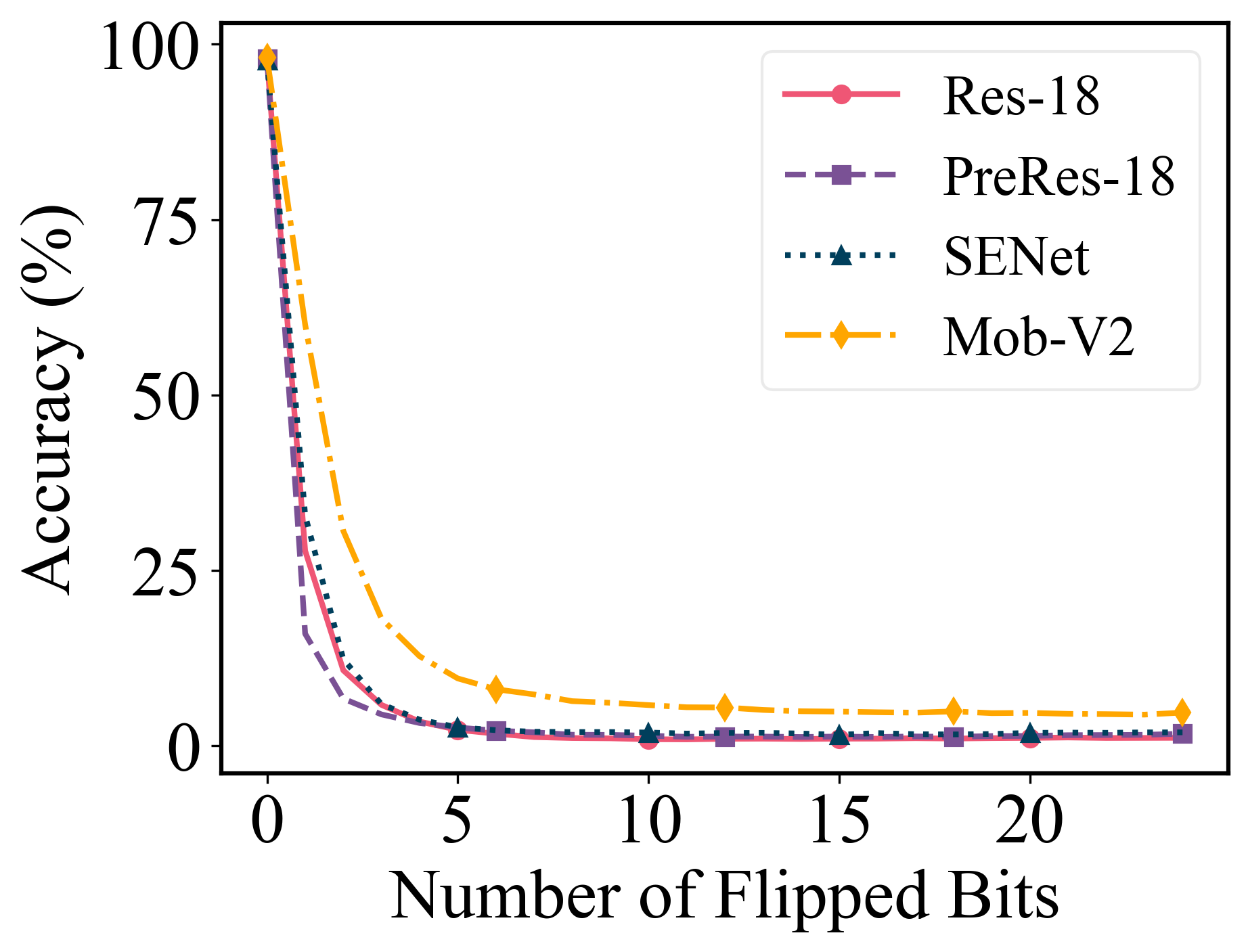}
                \caption{GTSRB}
        \end{subfigure}
        \begin{subfigure}{0.48\columnwidth}
                \includegraphics[width=\columnwidth]{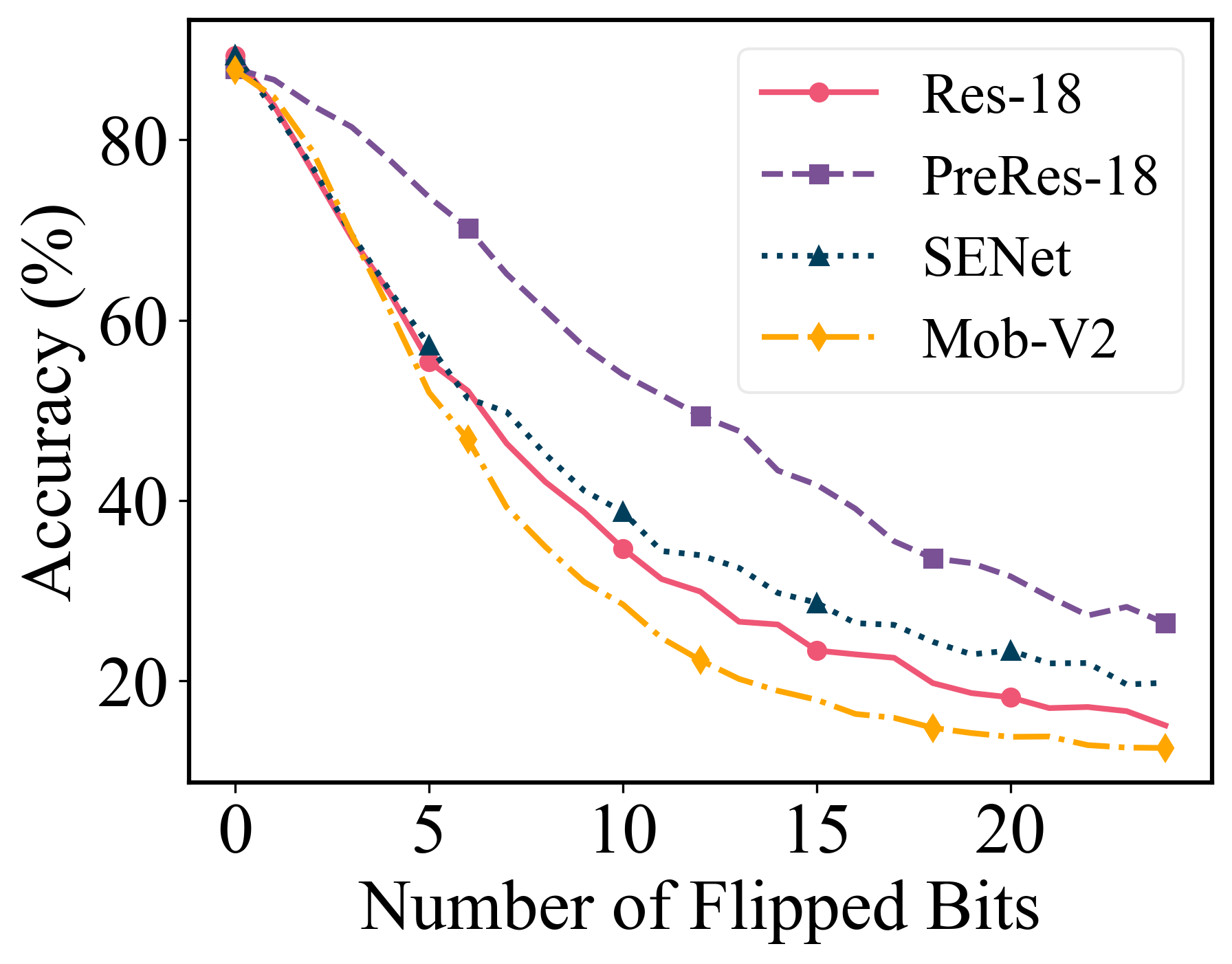}
                \caption{ImageNet-10}
        \end{subfigure}

        \caption{Accuracies of the protected models on test images embedded with incorrect keys with a few erroneous bits.}
        \vspace{-8mm}

        \label{fig:brute_force}
\end{figure}

\subsubsection{Uniqueness}

Ten protected models, each encoded with unique keys, are trained and evaluated across all possible model-key pair combinations. The confusion matrices are presented in Fig.~\ref{fig:confusion_matrices}, where each row represents a protected model and each column represents a key. The diagonal and off-diagonal values represent test accuracies for matched and mismatched model-key pairs, respectively. The off-diagonal values are consistently low (\textless20\%), indicating that user keys among protected models are non-exchangeable. This ensures precise ownership verification and culprit tracing as a protected model associated with one legitimate user cannot achieve accurate predictions on images encoded with another user's key.

An attacker may steal a protected model and the SteganoGAN encoder from a legitimate user, but lacks access to the correct key. To evaluate the tolerance of the protected model to bit errors in the key used for encoding the test images, we randomly flipped a few bits, ranging from 1 to 24, of the user keys. As shown in Fig.~\ref{fig:brute_force}, for the four low-resolution datasets, all protected models suffer from severe performance degradation with just one error bit. Although the accuracies do not drop to the level of random prediction, they are unacceptably low. When the number of flipped bits has increased to 3, the prediction accuracies of all protected models have dropped to no more than 21\%. On the other hand, for ImageNet-10, the accuracies of the models degraded slower with the number of flipped bits. The accuracies dropped by only around 4\% with just one bit flip in the user key. However, all four models experienced a performance degradation of more than 40\% when 13 key bits are flipped. Considering the 1024-bit key length, successfully brute-forcing remains computationally intractable.

\begin{table*}[t]
        \caption{DAs (\%) evaluated on an encoded dataset that contains 10,000 images embedded with unknown keys.} 
        \centering
        \renewcommand\arraystretch{1.1}
        \resizebox{0.9\linewidth}{!}{
                \begin{tabular}{c|cccccccccc|c}
                        \Xhline{2\arrayrulewidth}
                        \multirow{2}{*}{Dataset} & \multicolumn{10}{c|}{Key}                                                                                                                                                                                                                                                   &          \\ \cline{2-11} 
                                                 & \multicolumn{1}{c|}{$k_1$} & \multicolumn{1}{c|}{$k_2$} & \multicolumn{1}{c|}{$k_3$} & \multicolumn{1}{c|}{$k_4$} & \multicolumn{1}{c|}{$k_5$} & \multicolumn{1}{c|}{$k_6$} & \multicolumn{1}{c|}{$k_7$} & \multicolumn{1}{c|}{$k_8$} & \multicolumn{1}{c|}{$k_9$} & $k_{10}$ & Avg. DA \\ \Xhline{2\arrayrulewidth}
                        CIFAR-10                 & \multicolumn{1}{c|}{99.0}  & \multicolumn{1}{c|}{100.0} & \multicolumn{1}{c|}{100.0} & \multicolumn{1}{c|}{100.0} & \multicolumn{1}{c|}{100.0} & \multicolumn{1}{c|}{99.9}  & \multicolumn{1}{c|}{99.7}  & \multicolumn{1}{c|}{100.0} & \multicolumn{1}{c|}{99.9}  & 100.0  & 99.85    \\
                        Caltech-101              & \multicolumn{1}{c|}{97.9}  & \multicolumn{1}{c|}{99.2}  & \multicolumn{1}{c|}{99.3}  & \multicolumn{1}{c|}{99.1}  & \multicolumn{1}{c|}{99.0}  & \multicolumn{1}{c|}{98.8}  & \multicolumn{1}{c|}{98.8}  & \multicolumn{1}{c|}{99.4}  & \multicolumn{1}{c|}{99.4}  & 99.3   & 99.02    \\
                        GTSRB                    & \multicolumn{1}{c|}{99.7}  & \multicolumn{1}{c|}{99.8}  & \multicolumn{1}{c|}{100.0} & \multicolumn{1}{c|}{100.0} & \multicolumn{1}{c|}{100.0} & \multicolumn{1}{c|}{99.8}  & \multicolumn{1}{c|}{100.0} & \multicolumn{1}{c|}{100.0} & \multicolumn{1}{c|}{99.9}  & 99.7   & 99.89    \\
                        ImageNet-10              & \multicolumn{1}{c|}{100.0} & \multicolumn{1}{c|}{100.0} & \multicolumn{1}{c|}{100.0} & \multicolumn{1}{c|}{100.0} & \multicolumn{1}{c|}{100.0} & \multicolumn{1}{c|}{100.0} & \multicolumn{1}{c|}{99.9}  & \multicolumn{1}{c|}{100.0} & \multicolumn{1}{c|}{100.0} & 100.0  & 99.99    \\ \Xhline{2\arrayrulewidth}
                \end{tabular}
        }
        \label{tab:tracing_success_rate}
\end{table*}

\subsubsection{Licensee Verification}

For each dataset, we select 1,000 benign test images and encode them with 10 different keys to simulate an encoded dataset that contains 10,000 images embedded with unknown keys. These images are fed into the owner's private SteganoGAN decoder to extract their embedded keys. We set the threshold $\epsilon_3$ of Eq.~\eqref{eq:tsr} to 1, which is an extremely stringent threshold that allows only one bit error. Table~\ref{tab:tracing_success_rate} demonstrates that a high DA of over 97\% is achievable across all datasets and keys, with an average DA of above 99\% for all five datasets. This level of accuracy enables the owner or a trusted third party engaged by the owner to verify licensees with high confidence, even when provided with just a single encoded image by the user.

\begin{table*}[t]
        \caption{Comparison with existing active DNN IP protection methods on prediction performance (\%) on authorized and benign test datasets. $y^{\prime}=y+1$ is the wrong label of a clean image with ground truth label $y$. $b$ is the block size of the pixel-shuffling key. Ours$\ddagger$ denotes the elementary-distilled model that is trained with only the KL-divergence loss.}

        \centering
        \renewcommand\arraystretch{1.4}
        \resizebox{1.0\linewidth}{!}{
                \begin{tabular}{c|c|cc|cc|cc|cc}
                        \Xhline{2\arrayrulewidth}
                        \multirow{2}{*}{Method}          & Dataset $\rightarrow$    & \multicolumn{2}{c|}{CIFAR-10}                & \multicolumn{2}{c|}{Caltech-101}             & \multicolumn{2}{c|}{GTSRB}                  & \multicolumn{2}{c}{ImageNet-10}             \\ \cline{2-10} 
                                                         & Setting $\downarrow$     & \multicolumn{1}{c|}{Authorized}     & Benign & \multicolumn{1}{c|}{Authorized}     & Benign & \multicolumn{1}{c|}{Authorized}    & Benign & \multicolumn{1}{c|}{Authorized}    & Benign \\ \Xhline{2\arrayrulewidth}
                        DeepIPR~\cite{fan2021deepipr}    &     -                    & \multicolumn{1}{c|}{93.07 (-0.51)}  & 7.82   & \multicolumn{1}{c|}{74.89 (-0.84)}  & 4.30   & \multicolumn{1}{c|}{97.55 (+0.26)} & 11.75  & \multicolumn{1}{c|}{88.82 (-0.03)} & 38.96  \\ \hline
                        SSAT~\cite{xue2023ssat}          & $y^{\prime} = y + 1$     & \multicolumn{1}{c|}{93.53 (-0.05)}  & 0.60   & \multicolumn{1}{c|}{76.04 (+0.31)}  & 0.81   & \multicolumn{1}{c|}{97.85 (+0.56)} & 0.41   & \multicolumn{1}{c|}{87.57 (-1.28)} & 2.10   \\ \hline
                        DSN~\cite{tang2020deep}          &     -                    & \multicolumn{1}{c|}{93.06 (-0.52)}  & 11.55  & \multicolumn{1}{c|}{77.11 (+1.38)}  & 1.10   & \multicolumn{1}{c|}{97.97 (+0.68)} & 1.51   & \multicolumn{1}{c|}{89.04 (+0.19)} & 10.96  \\ \hline
                        \multirow{3}{*}{\begin{tabular}[c]{@{}c@{}}Pixel\\ Shuffling\\ \cite{pyone2020training}\end{tabular}}                                             & $b=2$                    & \multicolumn{1}{c|}{92.41 (-1.17)}  & 44.71  & \multicolumn{1}{c|}{73.48 (-2.25)}  & 41.27  & \multicolumn{1}{c|}{96.61 (-0.68)} & 8.52   & \multicolumn{1}{c|}{86.89 (-1.96)} & 58.45  \\
                                                         & $b=4$                    & \multicolumn{1}{c|}{89.28 (-4.30)}  & 23.78  & \multicolumn{1}{c|}{64.84 (-10.89)} & 17.46  & \multicolumn{1}{c|}{94.46 (-2.83)} & 8.80   & \multicolumn{1}{c|}{84.93 (-3.92)} & 56.50  \\ 
                                                         & $b=8$                    & \multicolumn{1}{c|}{82.36 (-11.22)} & 20.07  & \multicolumn{1}{c|}{53.14 (-22.59)} & 5.24   & \multicolumn{1}{c|}{88.19 (-9.10)} & 5.09   & \multicolumn{1}{c|}{83.37 (-5.48)} & 48.93  \\ \hline
                        Ours$\ddagger$                   & $\mathcal{L}_\text{kl}$      & \multicolumn{1}{c|}{92.84 (-0.74)}  & 9.90   & \multicolumn{1}{c|}{77.35 (+1.62)}  & 5.13   & \multicolumn{1}{c|}{98.23 (+0.94)} & 1.64   & \multicolumn{1}{c|}{89.88 (+1.03)} & 7.42   \\ \hline
                        Ours                             & $\mathcal{L}_\text{kl}+\mathcal{L}_\text{at}+\mathcal{L}_\text{crd}$ & \multicolumn{1}{c|}{93.15 (-0.43)}  & 10.13  & \multicolumn{1}{c|}{77.23 (+1.50)}  & 1.04   & \multicolumn{1}{c|}{97.78 (+0.49)} & 2.01   & \multicolumn{1}{c|}{89.31 (+0.46)} & 10.35  \\ \Xhline{2\arrayrulewidth}
                \end{tabular}
        }
        \label{tab:comparison_existing_works}
\end{table*}

\subsection{Active Protection Methods Comparison}

Four recent active protection methods, DeepIPR~\cite{fan2021deepipr}, SSAT~\cite{xue2023ssat}, DSN~\cite{tang2020deep}, and pixel shuffling~\cite{pyone2020training}, are re-implemented on ResNet-18 across the five datasets for comparison. We also directly distilled a student model from the MoE model with only KL-divergence and \textit{without} attention and contrastive representation loss terms. We call this model the ``elementary-distilled'' model. Table~\ref{tab:comparison_existing_works} presents the accuracies of the protected models on authorized and benign test images. The difference in accuracy between the authorized and benign test images for each protected model and dataset is also provided in bracket.

The performance of the pixel shuffling~\cite{pyone2020training} method is dependent on the shuffling block size $b$. A larger $b$ enhances effectiveness but compromises fidelity. For low-resolution datasets, $b = 2$ provides a good trade-off between effectiveness and fidelity. However, for ImageNet-10, it demonstrates poor performance on both effectiveness and fidelity. With $b = 8$, the prediction accuracy of the unauthorized images is approximately 50\%, while the accuracy of authorized images drops by over 5\%.

DeepIPR~\cite{fan2021deepipr}, SSAT~\cite{xue2023ssat}, and DSN~\cite{tang2020deep} can keep the accuracy drop to lower than 1.5\% on authorized images and sufficiently degrade the performance on unauthorized images (\textless40\%). However, these three methods have their own security vulnerability. The passport layers of DeepIPR can be easily localized since they require additional inputs. As demonstrated in~\cite{chen2023effective}, the attacker can replace these passport layers with custom normalization layers and re-train them to achieve high accuracy on benign images with a forged passport. SSAT assigns ground truth labels to poisoned training samples and wrong labels to clean training samples. The ground-truth label $y$ of clean input image is mapped to the wrong label $y^{\prime} = y+1$. This fixed bijective mapping can be easily reverse-engineered by observing the model's outputs to a small set of labeled test images. Any other deterministic bijective mapping is as vulnerable. The DSN method directly pastes a fixed binary pattern on the test images as the secret key, which can be easily spotted by human eyes.

IDEA achieves good effectiveness and fidelity on all datasets. Layer substitution attack is not able to break IDEA as we make no modification to the model architecture. Query-based reverse engineering is also infeasible as the outputs of unauthorized images are randomized. Since the perturbations introduced by SteganoGAN encoder are imperceivable and sample-specific, it is impossible for the attacker to recognize and copy the secret key.

Although the elementary-distilled model can achieve similar classification accuracy as the protected model, their differences will be more closely scrutinized in Sec.~\ref{sec:representation_analysis}.

\section{A Closer Look at Latent Representations} \label{sec:representation_analysis}

In this section, the hidden layer outputs of the unprotected, MoE, elementary-distilled and protected models are examined from three perspectives: MI estimation, attention map visualization, and representation visualization.

\begin{figure*}[t]
        \centering
        \begin{subfigure}{0.24\linewidth}
                \includegraphics[width=\linewidth]{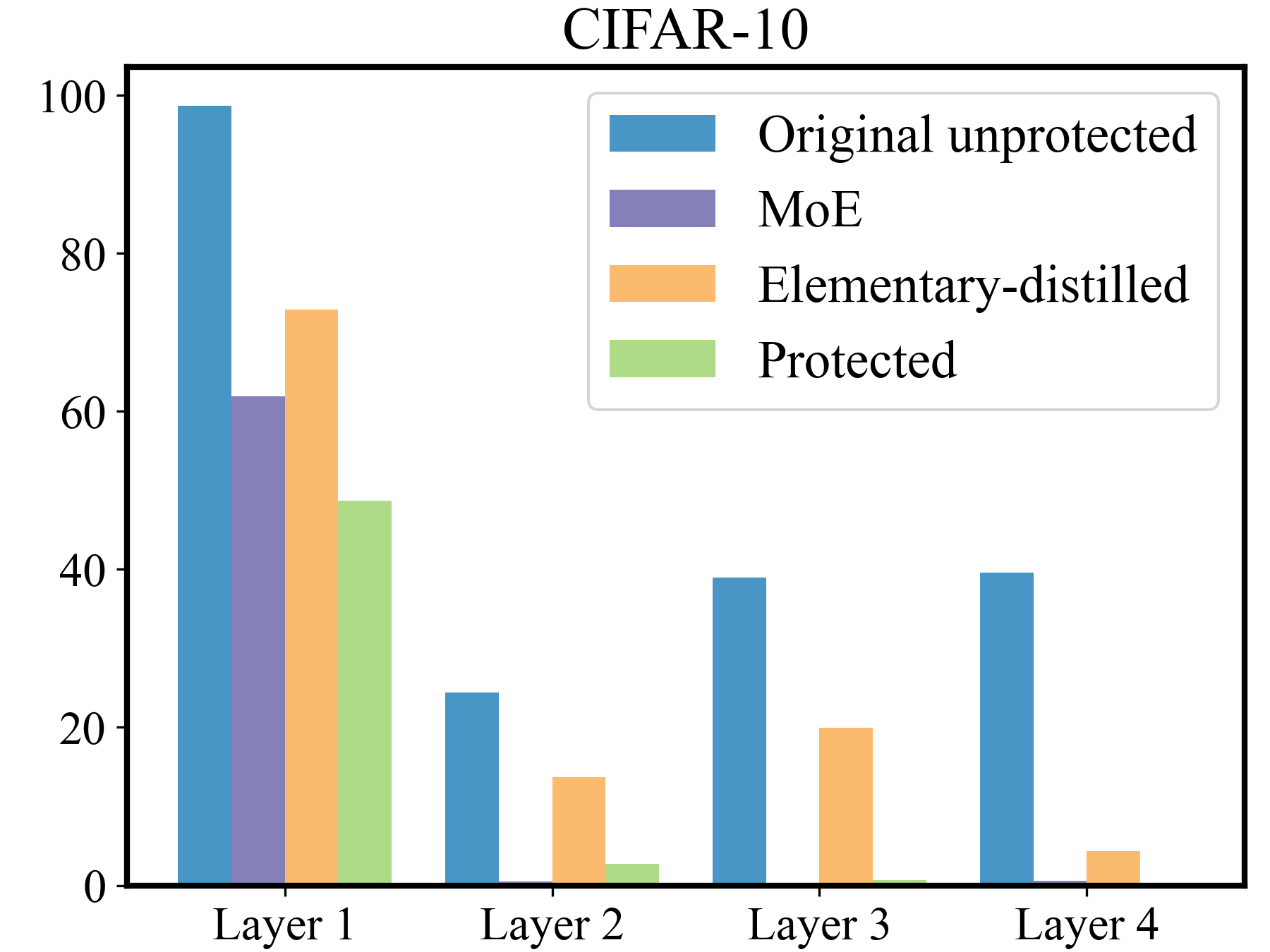}
        \end{subfigure}
        \begin{subfigure}{0.24\linewidth}
                \includegraphics[width=\linewidth]{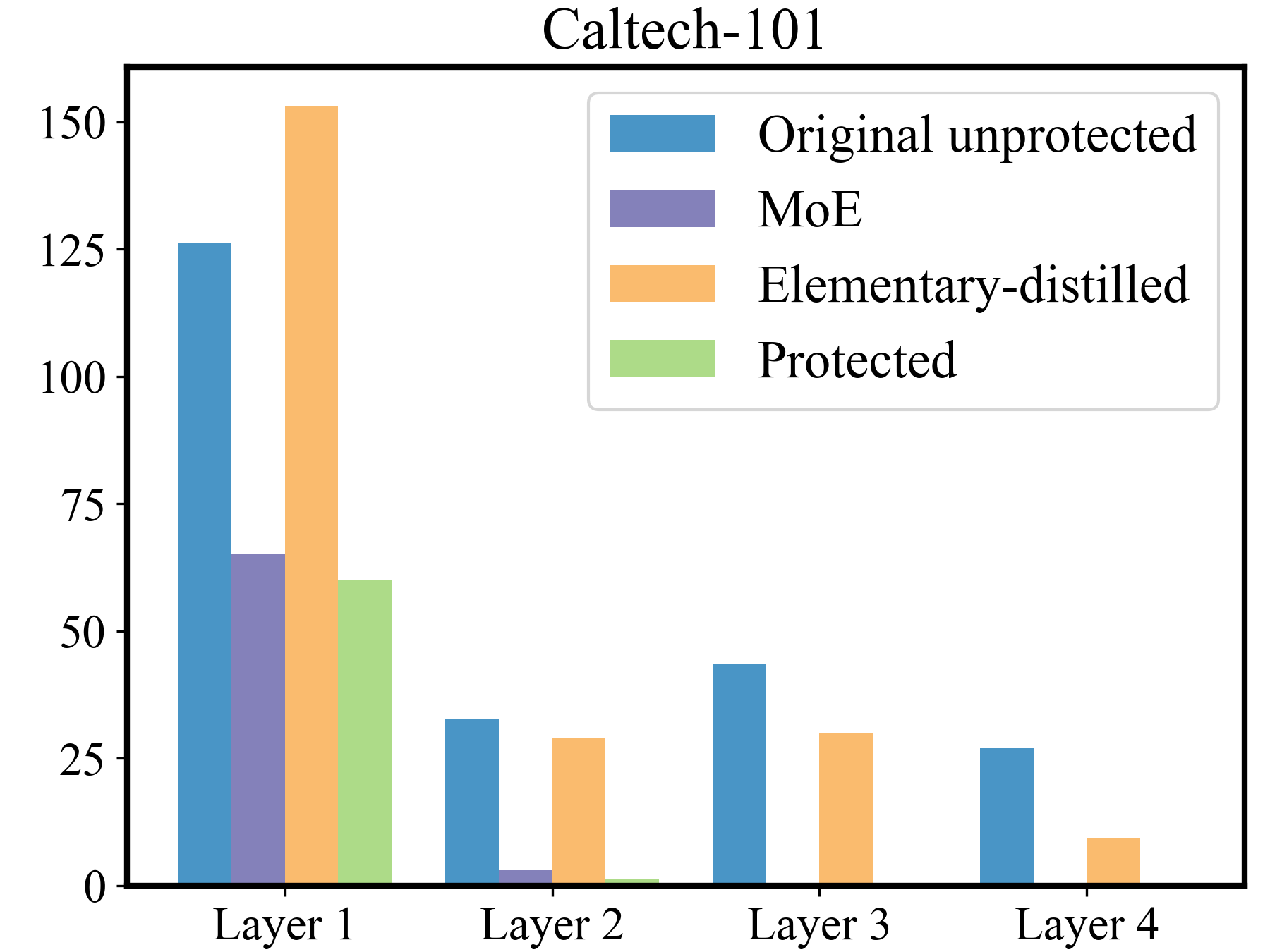}
        \end{subfigure}
        \begin{subfigure}{0.24\linewidth}
                \includegraphics[width=\linewidth]{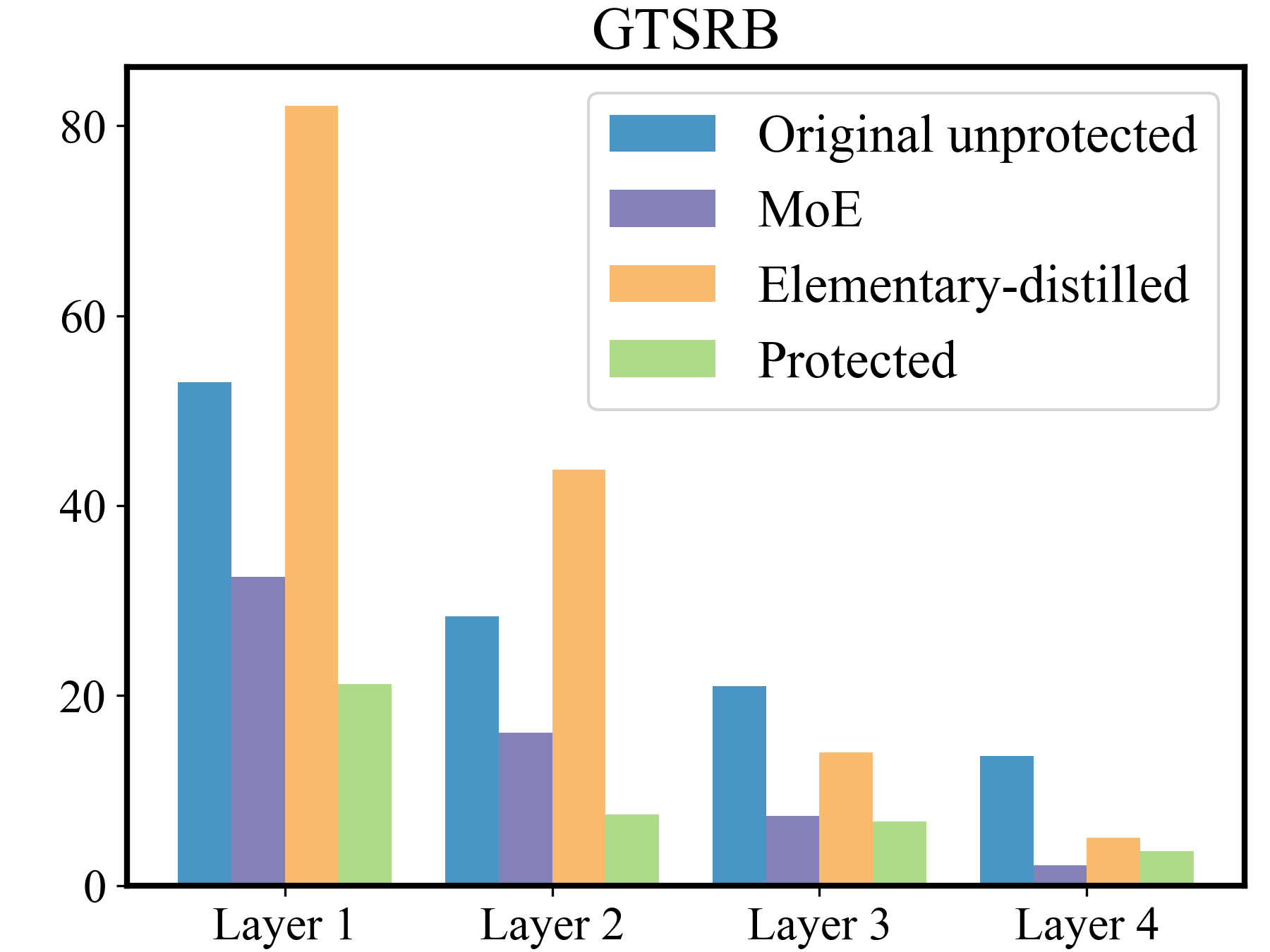}
        \end{subfigure}
        \begin{subfigure}{0.24\linewidth}
                \includegraphics[width=\linewidth]{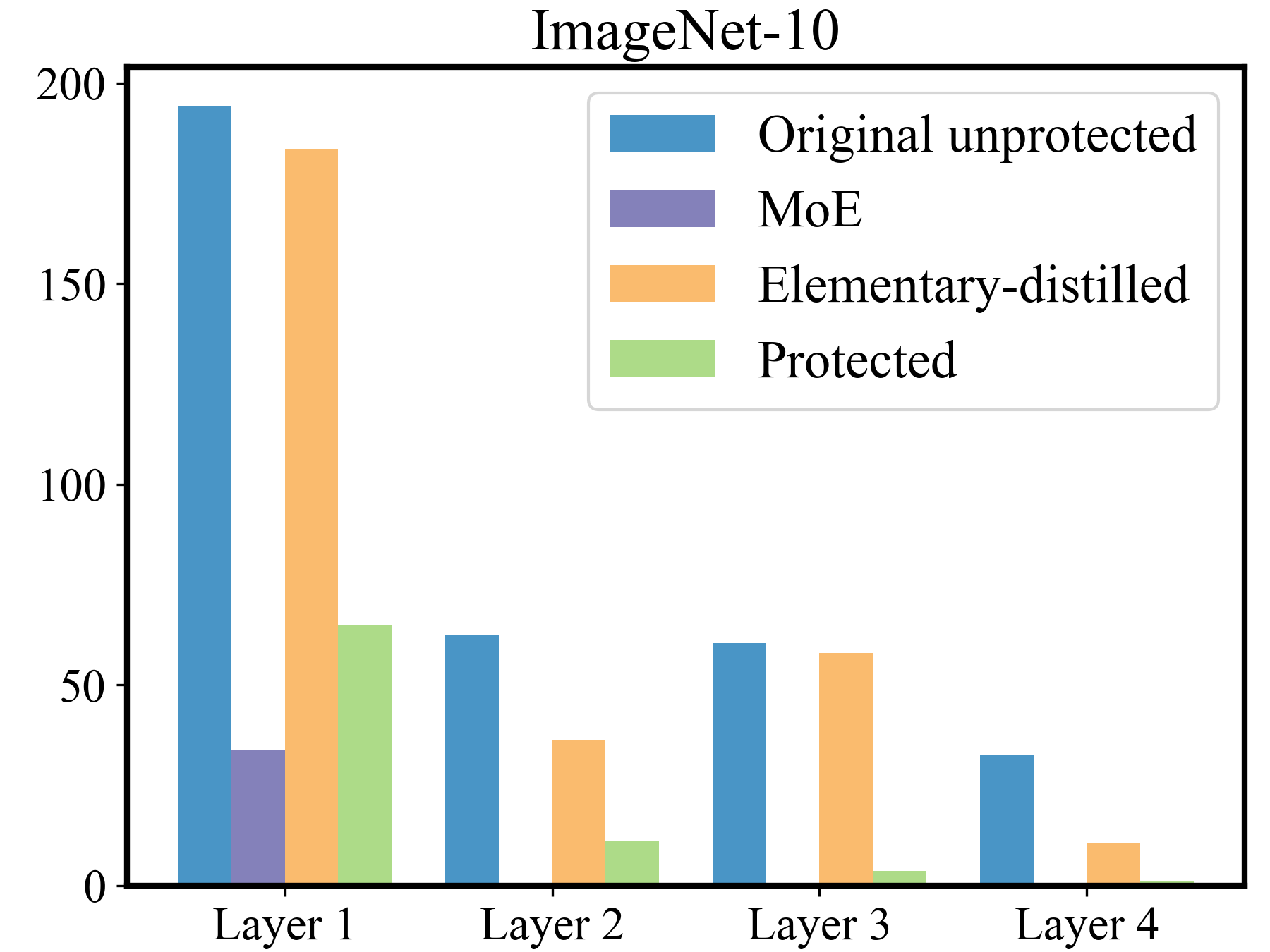}
        \end{subfigure}

        \vspace{1em}

        \begin{subfigure}{0.24\linewidth}
                \includegraphics[width=\linewidth]{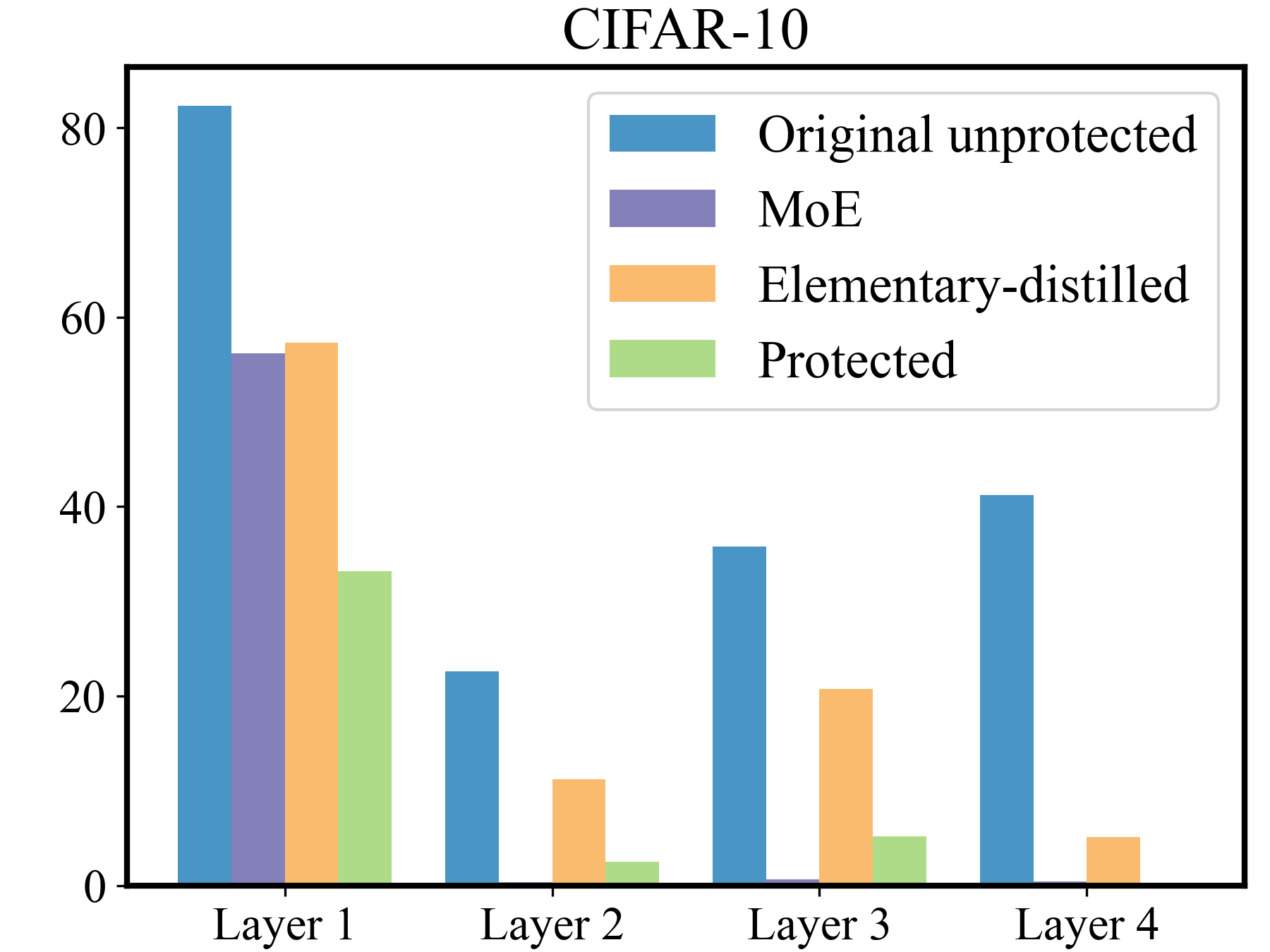}
        \end{subfigure}
        \begin{subfigure}{0.24\linewidth}
                \includegraphics[width=\linewidth]{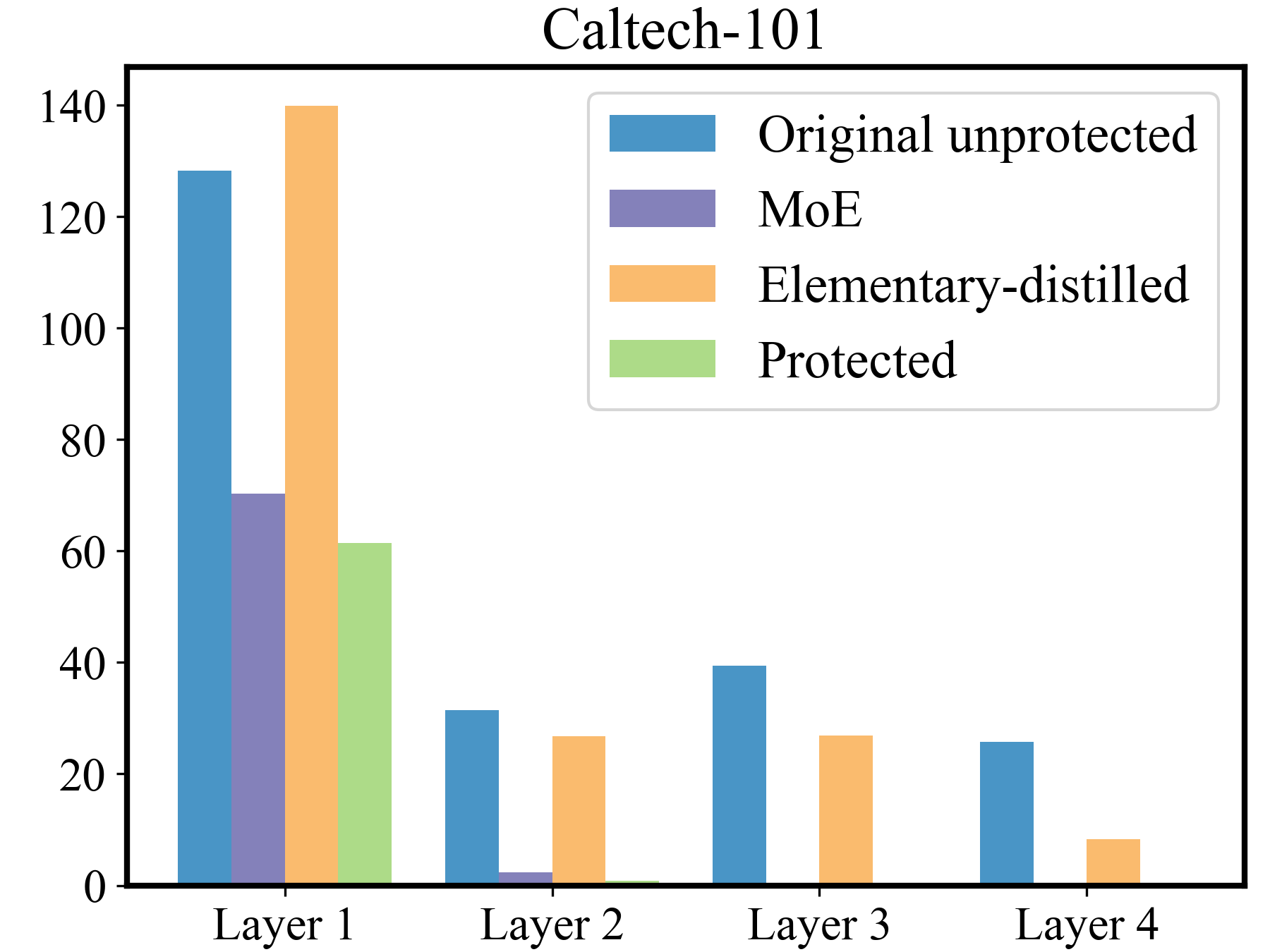}
        \end{subfigure}
        \begin{subfigure}{0.24\linewidth}
                \includegraphics[width=\linewidth]{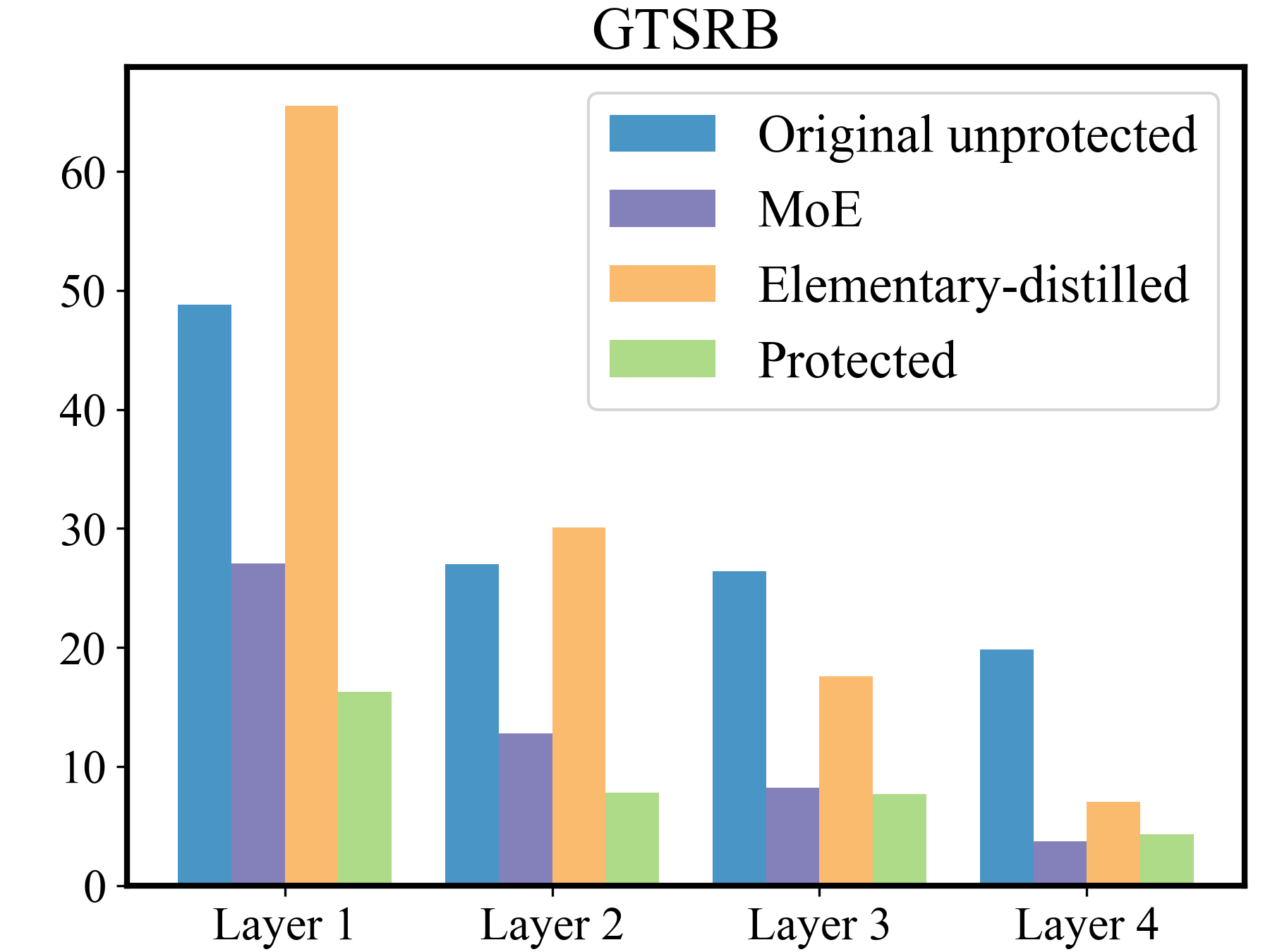}
        \end{subfigure}
        \begin{subfigure}{0.24\linewidth}
                \includegraphics[width=\linewidth]{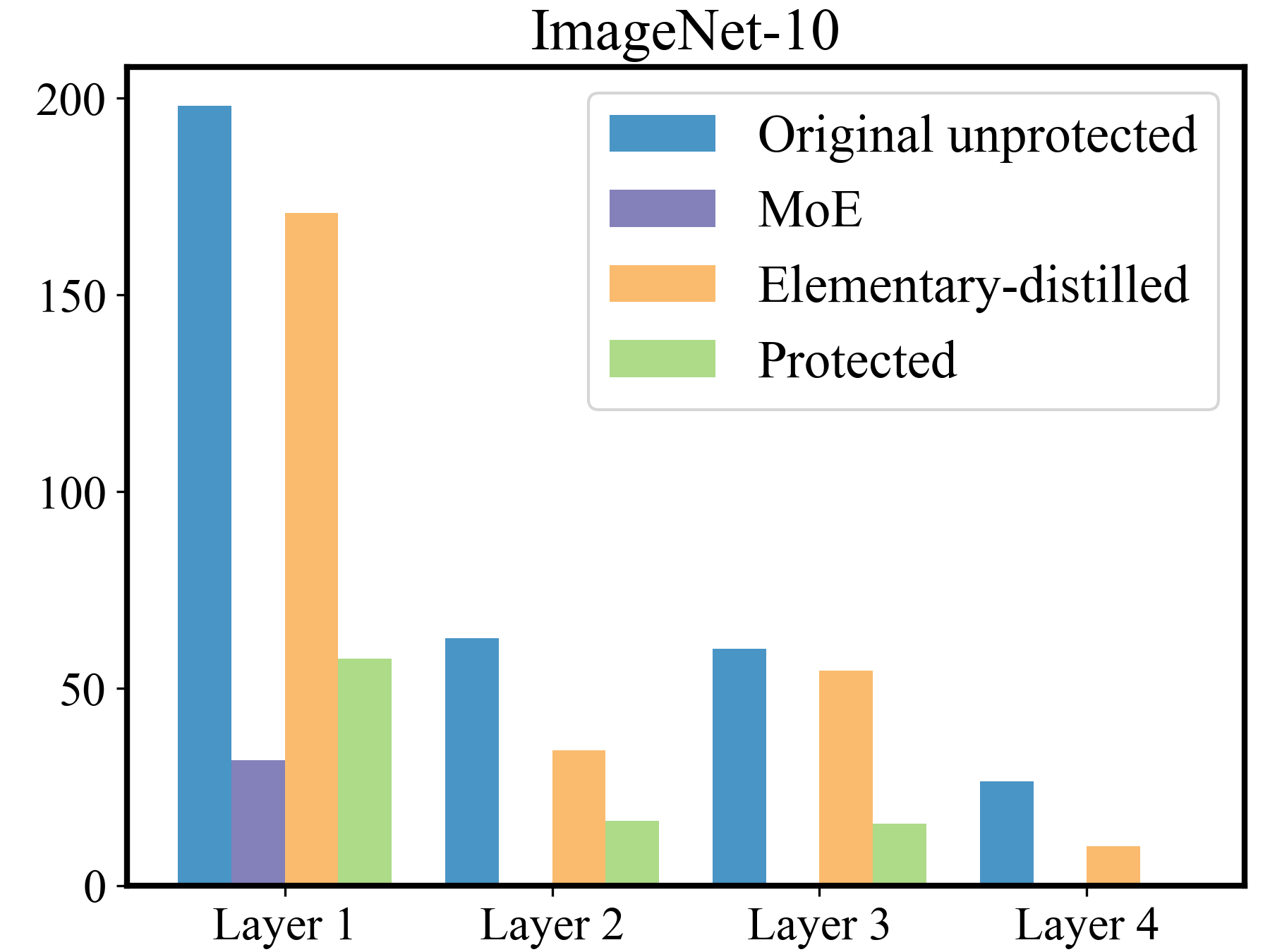}
        \end{subfigure}

        \caption{Estimated MI between the latent features of test images in the source (authorized) and target (unauthorized) domains. Due to the resolution of the vertical scale, some bars are not visible because their estimated MI values are very close to 0. Row 1: MI between paired authorized and benign images. Row 2: MI between paired authorized and noise images.}
        \vspace{-4mm}

        \label{fig:MI_estimation}
\end{figure*}

\subsection{MI Estimation}

Fig.~\ref{fig:MI_estimation} visualizes the estimated MI between the latent representations collected from paired source (authorized) and target (unauthorized) test samples, with the first row showing the MI between authorized and benign images, and the second row showing the MI between authorized and noise images. Layers 1 to 4 are the selected layers for MI minimization and multi-layer attention and contrastive representation knowledge distillation, progressing from the shallow to the deep layer.

For all selected layers, we observe that the estimated MI on the unprotected model is consistently high. This is expected, as this model has not been trained to recognize steganographic perturbations and thus produces similar representations for all types of images. In contrast, the MI between the source and target images estimated on the MoE model is significantly lower than that on the unprotected model, demonstrating that the training objectives in Eqs.~\eqref{eq:fake_expert_1} and~\eqref{eq:fake_expert_2} effectively increase the expected risk of the two fake experts. The protected model also shows similarly low MI values, indicating that the protected model faithfully emulates the MoE model's behavior across all selected layers. Consequently, if the input images are not embedded with the correct key, the adversary can only obtain meaningless intermediate layer representations that are comparable to those output by the fake experts. However, the MI estimated on the elementary-distilled model is higher than that of the MoE and protected models, and in some cases even higher than that of the unprotected model. This indicates that using only KL-divergence constraints on the output layer during knowledge distillation is insufficient to protect the model functionalities. Specifically, the shallow layers of the elementary-distilled model exhibit good generalizability on the target domains, making it easier to break the active authorization protection through fine-tuning, as these well-performed shallow layers require almost no adjustment.

\begin{figure*}[t]
        \centering
        \includegraphics[width=0.95\linewidth]{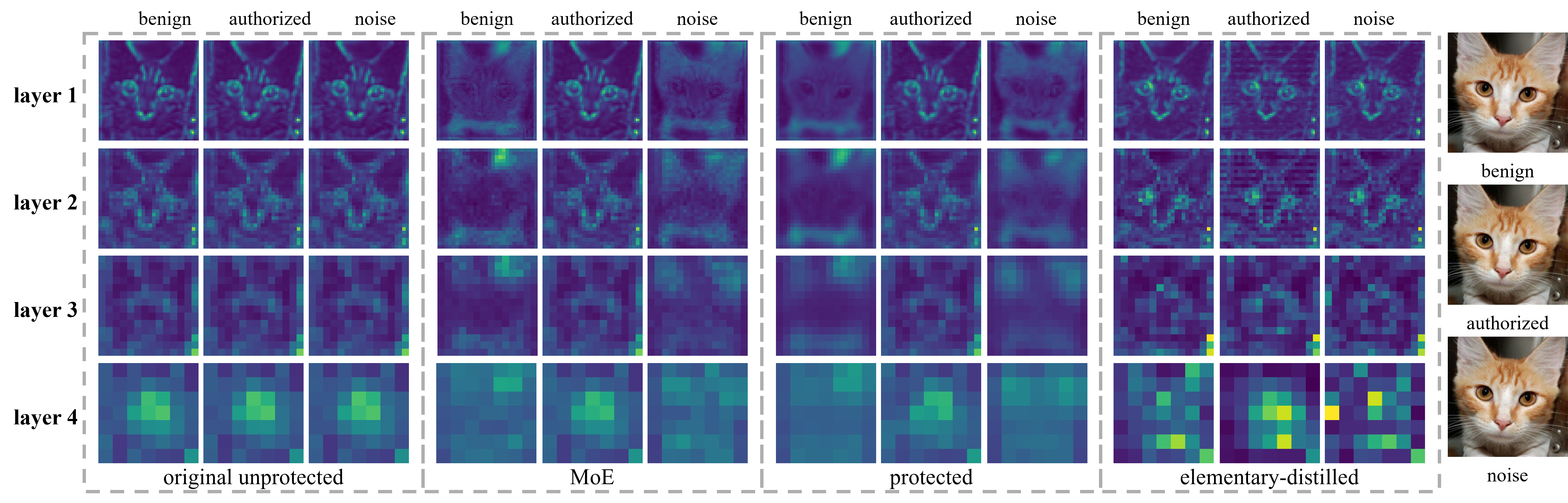}
        \caption{Visualization of attention maps generated by the original unprotected, MoE, protected and elementary-distilled models, respectively. The test images are randomly selected from ImageNet-10.}
        \label{fig:attention_map}
        \vspace{-2mm}
\end{figure*}

\subsection{Attention Map Visualization}

Fig.~\ref{fig:attention_map} shows the attention maps generated at the output of selected layers by feeding paired benign/authorized/noise test images into the four models. It shows that the unprotected model outputs similar attention maps for all three domains of test images at all selected layers. Conversely, the MoE model produces similar attention maps to the unprotected model only on input of authorized images. On input of benign or noise images, the MoE model exhibits attention behaviors akin to an untrained model to these inputs. Moreover, the protected model perfectly imitates the intermediate layer behaviors of the MoE model, as it outputs almost identical attention maps as those of the MoE model for the same domains of test images at all the selected layers. However, the elementary-distilled model exhibits similar attention maps for all types of images at the first three selected layers, and these attention maps are similar to those of the unprotected model at the same layers.

\begin{figure}[t]
        \centering
        \includegraphics[width=0.85\linewidth]{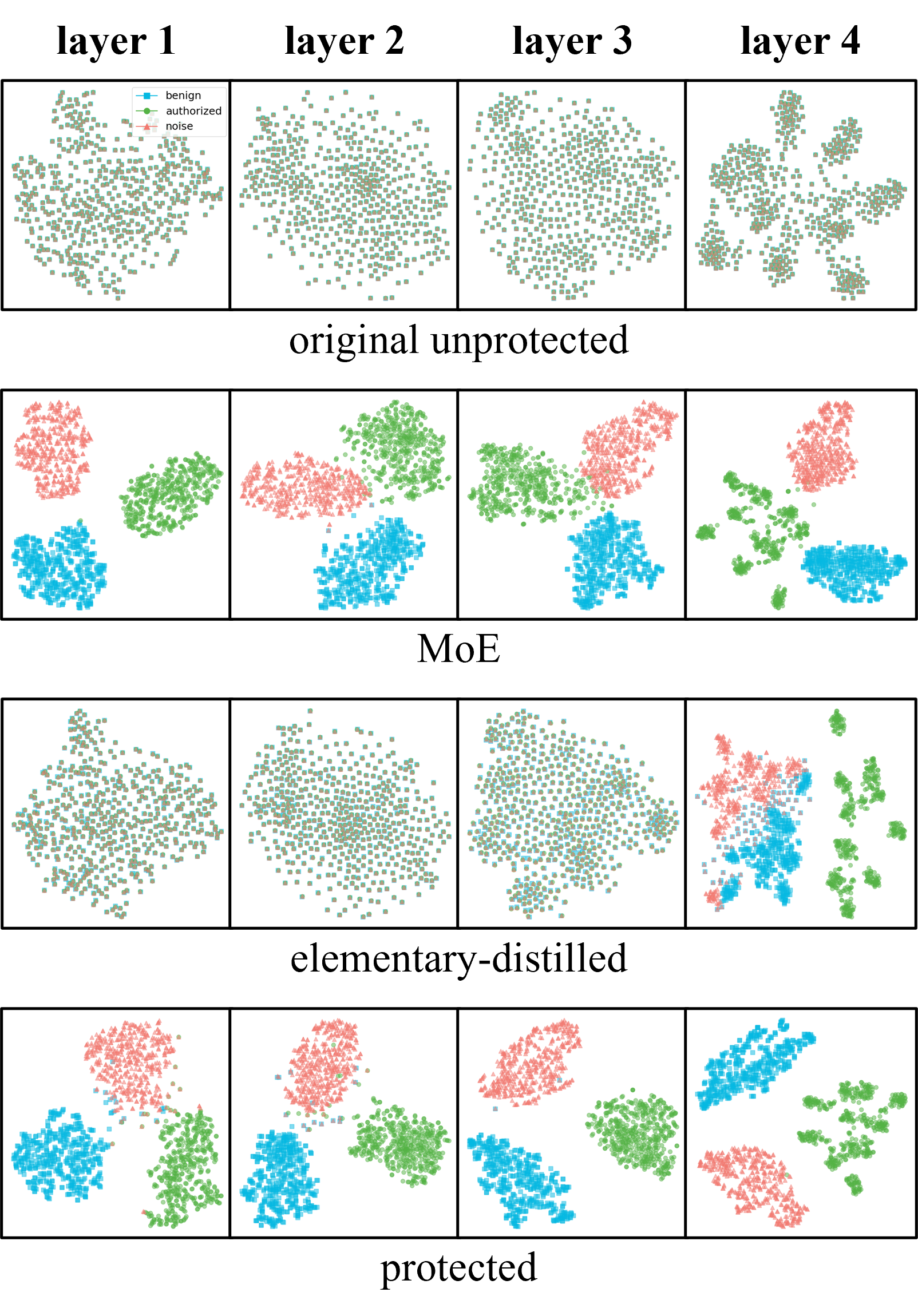}
        \caption{t-SNE visualization of the latent representations of the original unprotected, elementary-distilled, MoE and protected models obtained from 1,500 ImageNet-10 test images. The latent representations of the benign/authorized/noise test images are shown as \textbf{\textcolor[HTML]{05b9e2}{blue squares}}, \textbf{\textcolor[HTML]{54b345}{green dots}}, and \textbf{\textcolor[HTML]{f27970}{red triangles}}, respectively.}
        \label{fig:t-SNE_visualization}
\end{figure}

\subsection{t-SNE Visualization of Latent Representation}

We randomly selected 500 benign images from the ImageNet-10 dataset and generated their authorized and noise versions to create a mixed test set with 1,500 images. These test images were separately fed into the four models, and their outputs at the four selected layers were collected. The latent representations of the three types of test images are visualized using the t-SNE~\cite{van2008visualizing} in Fig.~\ref{fig:t-SNE_visualization}. For the unprotected model, the points corresponding to the paired benign/authorized/noise images almost completely overlap across all selected layers. This observation further illustrates that the source and target images have high estimated MI at the hidden layers. The elementary-distilled model has similar behaviors in layers 1 to 3, but the representations in layer 4 form three distinct clusters. This result, again, reveals that the elementary-distilled model has good generalizability at the shallow layers, and the representations of the three types of images are indistinguishable until the deep layers. On the contrary, for the MoE and protected models, the points corresponding to benign/authorized/noise images are well clustered across all selected layers. This indicates that the MI minimization has successfully made the two fake experts behave differently from the real expert, and the protected model effectively mimics the behavior of the MoE model at not only the output layer but also the selected hidden layers.

\section{Security Analysis}

We further investigate the robustness of IDEA against both model transformation and reverse engineering attacks. Model transformation attacks aim to remove the active authorization protection by directly modifying the pirated model through fine-tuning, model pruning, and transfer-learning. reverse engineering attacks aim to reverse engineer the encoding process without modifying the pirated model.

\begin{figure}[t]
        \centering
        \begin{subfigure}{0.49\columnwidth}
                \includegraphics[width=\columnwidth]{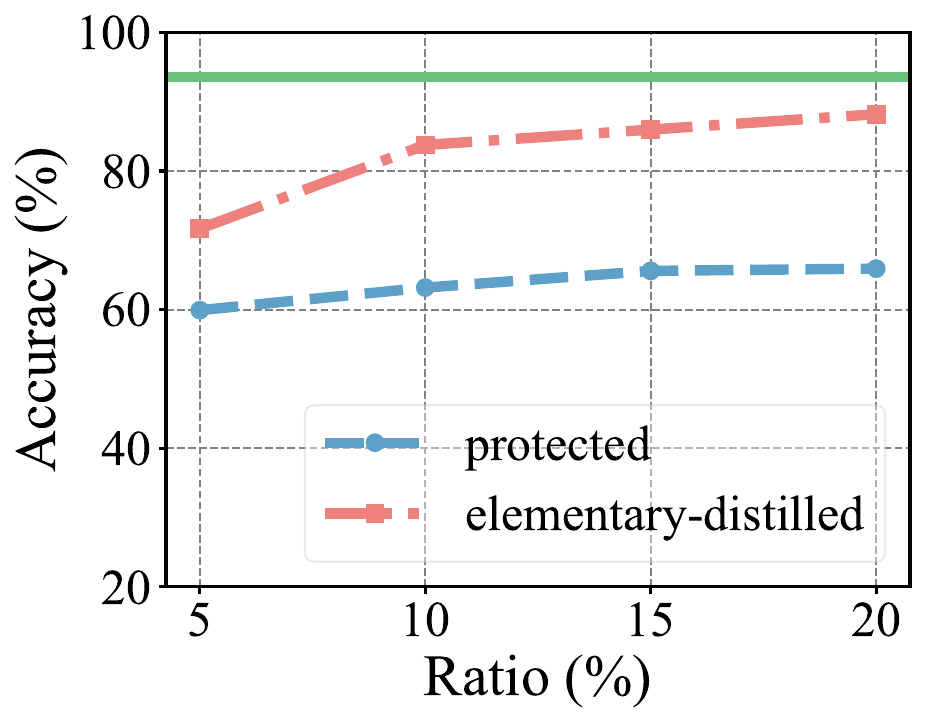}
                \caption{CIFAR-10}
        \end{subfigure}
        \begin{subfigure}{0.48\columnwidth}
                \includegraphics[width=\columnwidth]{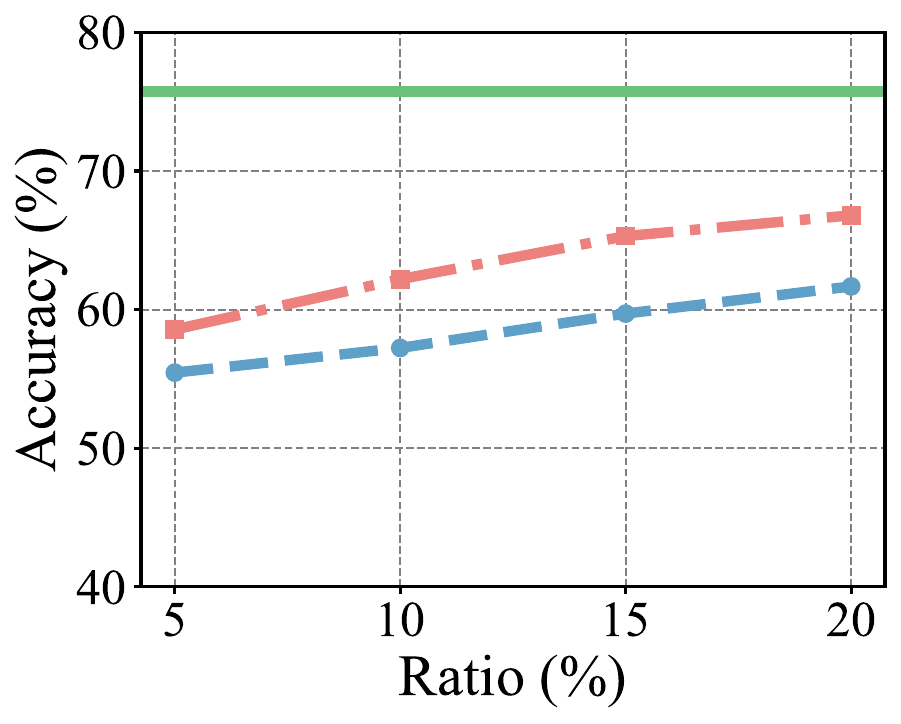}
                \caption{Caltech-101}
        \end{subfigure}
        \begin{subfigure}{0.49\columnwidth}
                \includegraphics[width=\columnwidth]{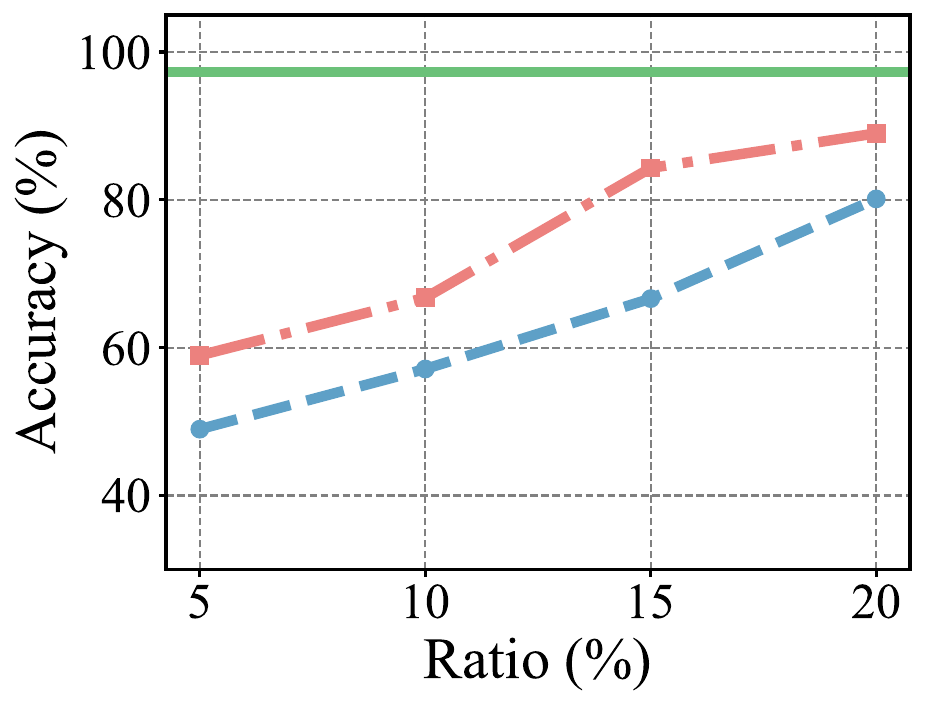}
                \caption{GTSRB}
        \end{subfigure}
        \begin{subfigure}{0.48\columnwidth}
                \includegraphics[width=\columnwidth]{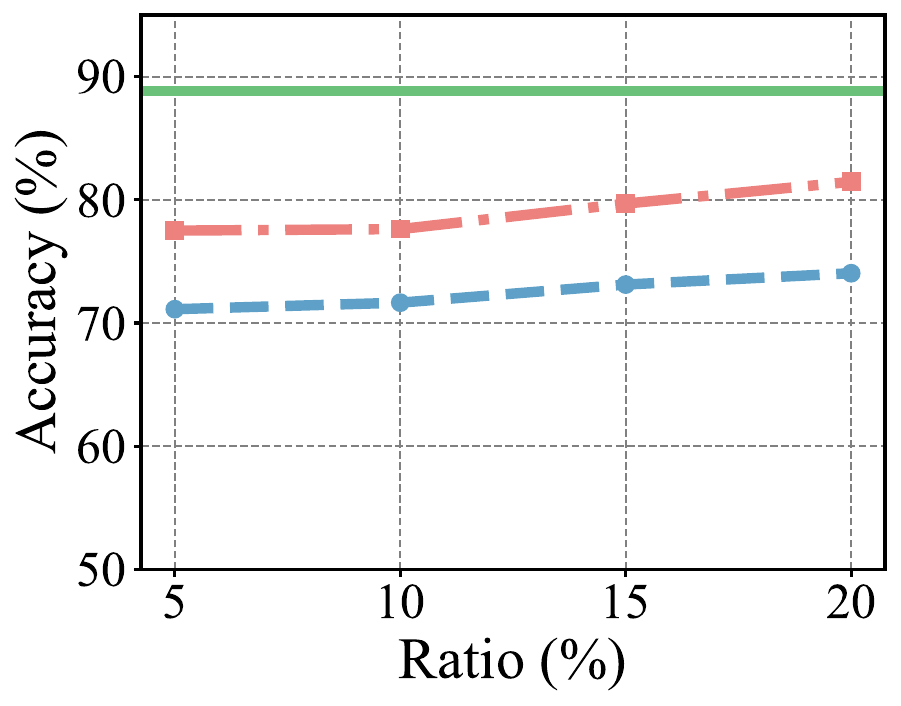}
                \caption{ImageNet-10}
        \end{subfigure}

        \caption{Performance of fine-tuned models on benign test datasets. The green solid lines indicate the baseline performance of unprotected models.}

        \label{fig:fine_tuning}
\end{figure}

\subsection{Model Transformation Attacks}

\subsubsection{Fine-tuning} An attacker may fine-tune a stolen model to bypass active control. It is assumed that the attacker has access to only limited training data compared to the owner; otherwise, the attacker could simply train their own model from scratch.

To evaluate the robustness of IDEA against such an attack, both protected and elementary-distilled models were fine-tuned for 100 epochs using 5\%, 10\%, 15\%, and 20\% samples of the benign training dataset. Fig.~\ref{fig:fine_tuning} illustrates the accuracy of the fine-tuned models on the benign test dataset. The baseline performance of the unprotected models is represented by the green solid line. For fine-tuned protected models, the performance improves as the ratio of available samples increases. However, even with 20\% training samples, the inference accuracy on benign data remains significantly lower than the baseline performance. Additionally, models fine-tuned from the elementary-distilled version exhibit higher accuracy than those fine-tuned from the protected model. This observation aligns with the analysis in Sec.~\ref{sec:representation_analysis} that the shallow layers of the elementary-distilled model are more vulnerable to exploitation during fine-tuning attacks. To mitigate this, multi-layer attention and contrastive representation losses are necessary to prevent the shallow layers of the distilled model from leaking the real expert's functionality.

\begin{table*}[t]
        \caption{Accuracy (\%) of models trained using transfer-learning from unprotected and protected models. The original training datasets are listed in the first row and the private dataset to be adapted in the first column. ImageNet-10$\dagger$ is another subset of ImageNet consisting of 10 classes that do not overlap with ImageNet-10.}

        \centering
        \renewcommand\arraystretch{1.1}
        \resizebox{0.9\linewidth}{!}{
                \begin{tabular}{c|cc|cc|cc|cc}
                        \hline
                        \multirow{2}{*}{Dataset} & \multicolumn{2}{c|}{CIFAR-10}                & \multicolumn{2}{c|}{Caltech-101}             & \multicolumn{2}{c|}{GTSRB}                   & \multicolumn{2}{c}{ImageNet-10}              \\ \cline{2-9} 
                                                 & \multicolumn{1}{c|}{unprotected} & protected & \multicolumn{1}{c|}{unprotected} & protected & \multicolumn{1}{c|}{unprotected} & protected & \multicolumn{1}{c|}{unprotected} & protected \\ \hline
                        CIFAR-10                 & \multicolumn{1}{c|}{-}           & -         & \multicolumn{1}{c|}{37.35}       & 26.05     & \multicolumn{1}{c|}{50.85}       & 22.80     & \multicolumn{1}{c|}{-}           & -         \\
                        Caltech-101              & \multicolumn{1}{c|}{68.17}       & 54.62     & \multicolumn{1}{c|}{-}           & -         & \multicolumn{1}{c|}{63.06}       & 43.95     & \multicolumn{1}{c|}{-}           & -         \\
                        GTSRB                    & \multicolumn{1}{c|}{46.78}       & 35.43     & \multicolumn{1}{c|}{31.99}       & 20.00     & \multicolumn{1}{c|}{-}           & -         & \multicolumn{1}{c|}{-}           & -         \\
                        ImageNet-10              & \multicolumn{1}{c|}{-}           & -         & \multicolumn{1}{c|}{-}           & -         & \multicolumn{1}{c|}{-}           & -         & \multicolumn{1}{c|}{50.34}       & 40.36     \\ \hline
                \end{tabular}
        }
        \label{tab:transfer_learning}
\end{table*}

\begin{figure}[t]
        \centering
        \begin{subfigure}{0.48\linewidth}
            \includegraphics[width=\textwidth]{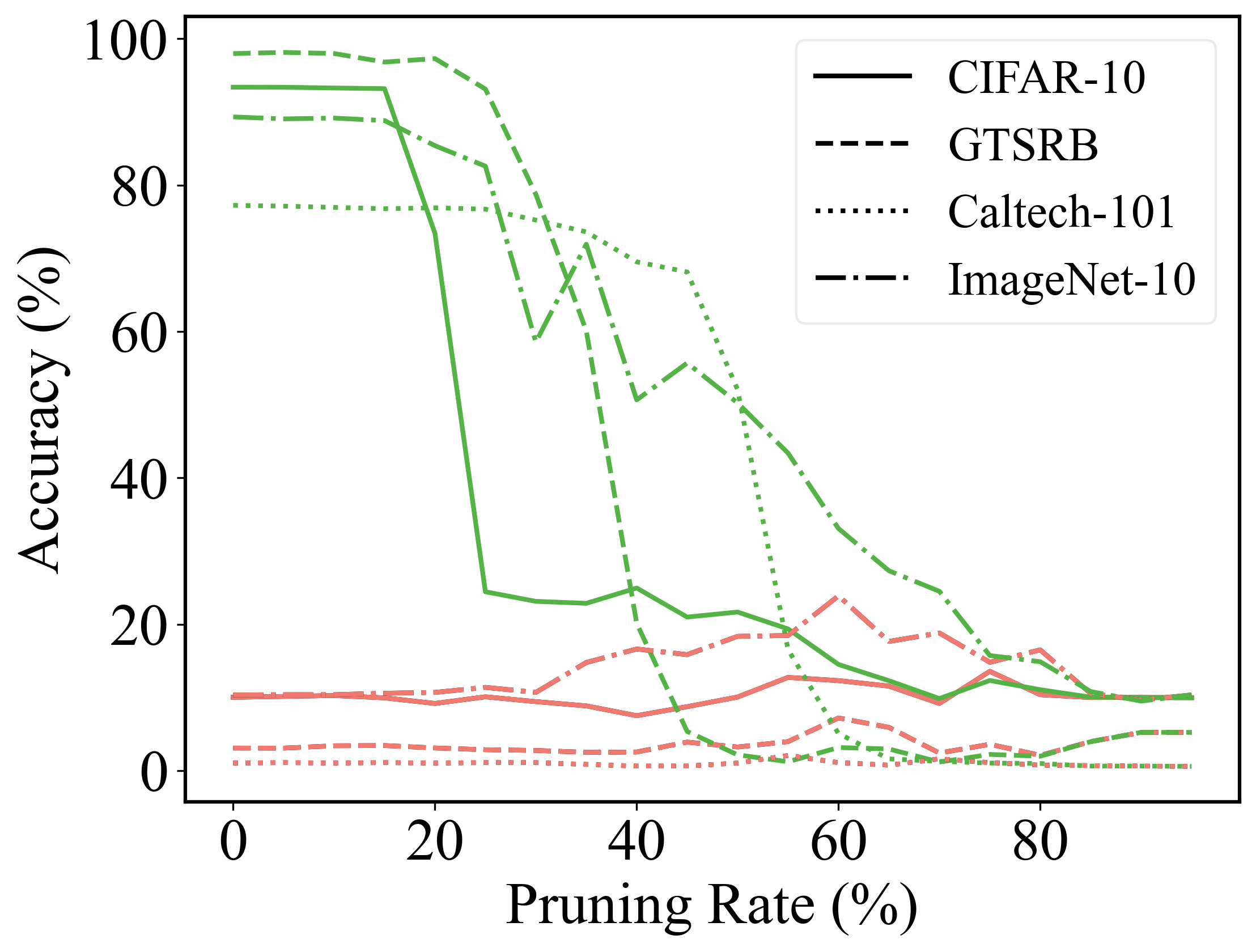}
            \caption{weight-pruning}
            \label{fig:weight_pruning}
        \end{subfigure}
        \begin{subfigure}{0.48\linewidth}
            \includegraphics[width=\textwidth]{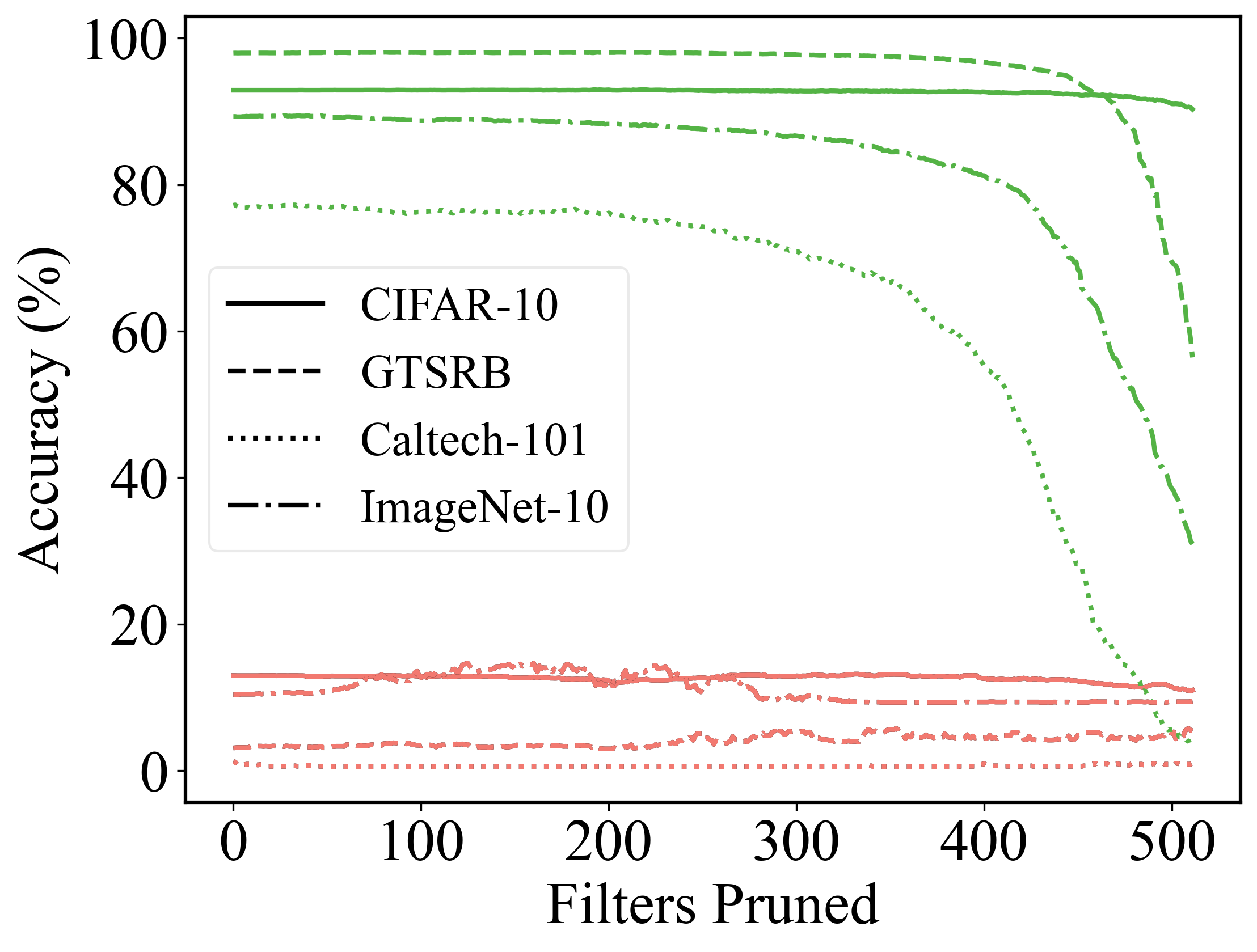}
            \caption{filter-pruning}
            \label{fig:filter_pruning}
        \end{subfigure}

        \caption{Robustness of IDEA against weight-pruning and filter-pruning attacks. We present the pruned models' performance on authorized and benign (unauthorized) test datasets in \textbf{\textcolor[HTML]{54b345}{green}} and \textbf{\textcolor[HTML]{f27970}{red}}, respectively.}

        \label{fig:model_pruning}
\end{figure}

\subsubsection{Model pruning} Weight pruning (WP) and filter pruning (FP), are commonly used to reduce model size by removing parameters that contribute insignificantly to inference performance~\cite{han2015learning}. WP globally prunes a fraction (ranging from 5\% to 95\%, with a step size of 5\%) of the smallest magnitude weights. FP iteratively prunes filters in a specific convolutional layer, starting from the smallest magnitude to the largest. Following the configuration in~\cite{liu2018fine}, we select the last convolutional layer of ResNet-18 as the target layer for FP implementation.

The results are shown in Fig.~\ref{fig:model_pruning}. For WP, the accuracy of the protected model on both test images with and without the correct user key embedded changes minimally when the pruning rate is less than 30\%. As the pruning rate increases, the inference performance of test images embedded with correct user key drops sharply due to the removal of some crucial weights. Nevertheless, the accuracy on benign test images remains consistently low across all pruning rates. Similarly, the accuracy on benign test images falls below 20\% consistently across all datasets regardless of the number of filters pruned by the FP attack. As the functionalities of real and fake experts are tightly coupled by knowledge distillation, it is infeasible to disable the active authorization by merely pruning weights or filters without sacrificing the inference accuracy.

\subsubsection{Transfer-learning} This attack aims to utilize the well-trained feature extractor to adapt the pirated model to a different task using a private dataset. The common transfer-learning paradigm freezes all layers except the last fully connected layer, which is randomly initialized based on the new task. For the four low-resolution datasets, we perform transfer learning on both unprotected and protected models using other low-resolution datasets. For the high-resolution ImageNet-10, we select another 10 classes from the entire ImageNet dataset to form a non-overlapping subset ImageNet-10$\dagger$ for transfer-learning. The results in Tabel~\ref{tab:transfer_learning} show that models that are transfer-learnt from protected models exhibit worse performance than from unprotected models as protected models output meaningless features by imitating fake experts.

\begin{table}[t]
        \caption{Accuracy (\%) of pirated models on reverse-engineered images generated based on Assumptions 1 and 2.}

        \centering
        \renewcommand\arraystretch{1.1}
        \resizebox{0.9\linewidth}{!}{
                \begin{tabular}{c|c|ccccc}
                        \Xhline{2\arrayrulewidth}
                        \multirow{2}{*}{Dataset}     & \multirow{2}{*}{Assumption} & \multicolumn{5}{c}{\# Available images / pairs}    \\ \cline{3-7} 
                                                     &                             & \multicolumn{1}{c|}{500}   & \multicolumn{1}{c|}{1000}  & \multicolumn{1}{c|}{2000}  & \multicolumn{1}{c|}{3000}  & 5000  \\ \Xhline{2\arrayrulewidth}
                        \multirow{2}{*}{CIFAR-10}    & 1                           & \multicolumn{1}{c|}{9.67}  & \multicolumn{1}{c|}{12.23} & \multicolumn{1}{c|}{12.12} & \multicolumn{1}{c|}{10.06} & 10.00 \\ 
                                                     & 2                           & \multicolumn{1}{c|}{10.00} & \multicolumn{1}{c|}{12.71} & \multicolumn{1}{c|}{13.39} & \multicolumn{1}{c|}{16.06} & 56.84 \\ \hline
                        \multirow{2}{*}{Caltech-101} & 1                           & \multicolumn{1}{c|}{0.98}  & \multicolumn{1}{c|}{1.21}  & \multicolumn{1}{c|}{8.47}  & \multicolumn{1}{c|}{1.21}  & 22.13 \\  
                                                     & 2                           & \multicolumn{1}{c|}{1.04}  & \multicolumn{1}{c|}{5.53}  & \multicolumn{1}{c|}{22.71} & \multicolumn{1}{c|}{29.80} & 37.46 \\ \hline
                        \multirow{2}{*}{GTSRB}       & 1                           & \multicolumn{1}{c|}{4.99}  & \multicolumn{1}{c|}{4.96}  & \multicolumn{1}{c|}{5.70}  & \multicolumn{1}{c|}{4.99}  & 0.56  \\  
                                                     & 2                           & \multicolumn{1}{c|}{4.05}  & \multicolumn{1}{c|}{3.52}  & \multicolumn{1}{c|}{30.11} & \multicolumn{1}{c|}{34.70} & 44.51 \\ \hline
                        \multirow{2}{*}{ImageNet-10} & 1                           & \multicolumn{1}{c|}{9.65}  & \multicolumn{1}{c|}{10.19} & \multicolumn{1}{c|}{16.15} & \multicolumn{1}{c|}{14.38} & 15.92 \\  
                                                     & 2                           & \multicolumn{1}{c|}{10.23} & \multicolumn{1}{c|}{10.92} & \multicolumn{1}{c|}{15.58} & \multicolumn{1}{c|}{16.19} & 27.85 \\ \Xhline{2\arrayrulewidth}
                \end{tabular}
        }
        \label{tab:reverse_engineering}
\end{table}

\subsection{Reverse engineering Attacks}

This kind of attacks assumes that the adversary has successfully stolen a protected model from a legitimate user, but not the SteganoGAN encoder and the correct user key. We make two assumptions about the data available to the attacker.

\textit{Assumption 1:} The attacker only has a small portion of the benign images to train a surrogate generator to produce encoded images for the pirated model.

\textit{Assumption 2:} The attacker has not only a small portion of benign images but also their authorized versions embedded with the correct key by eavesdropping on legitimate users. This scenario enables the attacker to train the generator more efficiently by simultaneously minimizing the MSE loss between the generated and authorized images.

We simulate these attacks by training a generator with different numbers of benign images or benign-authorized pairs, ranging from 500 to 5,000. The generator has a similar architecture to the SteganoGAN encoder but without the key input port. Both its input and output are 3-channel color images of the same size. Table~\ref{tab:reverse_engineering} reports the accuracy of the pirated model on test images created by the generator. Under \textit{Assumption 1}, most reverse engineering attacks result in low prediction accuracy close to random guessing. The best attack performance achieved on Caltech-101 with 5,000 available benign images is merely 22.13\%. As expected, the attack performance under \textit{Assumption 2} is much better in most cases, with an average improvement of 9.55\%. However, even with 5,000 available benign-authorized pairs, which are difficult for the attacker to collect in reality, the best attack performance of 56.84\% achievable on CIFAR-10 is still far lower than the baseline. Therefore, IDEA is resilient against even the strongest assumption of reverse engineering attacks.

\section{Limitations}
\label{sec:limitations}
Security comes at a cost. Notwithstanding the robust proactive protection for DNNs offered by IDEA, there are inevitable trade-offs.

One premium to pay for the active control is the computational cost of generating protected models. Each instance requires training three experts and performing knowledge distillation to integrate their functionalities into a unified network. This limitation is similarly observed in existing active IP protection mechanisms~\cite{ren2022protecting, xue2023ssat, pyone2020training, tang2020deep, fan2021deepipr, zhou2023nnsplitter, li2024securenet} designed for multiple users. Nonetheless, this process is faster than training a model from scratch, and the overhead remains manageable. For the distributed revenue business model, this cost adds only incrementally to the bracket price to purchase a customer-premise model for unlimited private use. Another aspect is the number of queries required for ownership verification, which scales linearly with the number of users. Even with a small query dataset, such as 100 samples, encoding these with all recorded keys introduces additional computational steps. However, this is a one-time process for verifying ownership and identifying dishonest users. The overhead is trivial relative to the strong protection it offers. Lastly, encoding test images through the SteganoGAN encoder introduces a slight latency during offline usage.

\section{Conclusion}

We have presented a novel active DNN IP protection method called IDEA. It steganographically encodes distributed model instances with user-specific keys. The authorized user of a protected model instance can unlock its inference performance by submitting test images embedded with a valid key assigned by the model owner. The licensee can be verified through SteganoGAN decoding with almost 100\% accuracy, without exposing authorized users' secret keys. In the event of IP infringement, the owner can verify the authenticity of deployed models and trace the culprit by querying the suspected model with a small set of test samples or decoding the user submitted queries. Extensive experiments conducted across four image classification datasets and four DNN models validated the excellent inference performance on authorized inputs, and corroborated that IDEA-protected models cannot be unlocked by images encoded with incorrect keys even with only a few flipped bits or keys stolen from other legitimate users. This strong uniqueness property is important to assure that the tracking of any redistributed models to its culprit is non-repudiable. IDEA is also robust against model transformation and reverse engineering attacks, and more stealthy, effective, and scalable with networks and dataset sizes compared with existing active protection methods.

\bibliographystyle{IEEEtran}
\bibliography{main}


 





\end{document}